\documentclass[%
 aip,
 amsmath,amssymb,
 reprint,%
]{revtex4-1}

\usepackage{graphicx}
\usepackage{cancel}
\usepackage{empheq}
\usepackage{dcolumn}
\usepackage{bm}
\usepackage{xcolor}

\begin{document}

\preprint{APS/123-QED}

\title{Numerical Study of Magnetic Island Coalescence Using Magnetohydrodynamics With Adaptively Embedded Particle-In-Cell Model}

\author{Dion Li}
\affiliation{Department of Nuclear Engineering and Radiological Sciences, University of Michigan, Ann Arbor, Michigan 48109, USA}%

\author{Yuxi Chen}
\affiliation{Department of Astrophysical Sciences, Princeton University, Princeton, New Jersey 08544, USA}%
\affiliation{Princeton Plasma Physics Laboratory, Princeton University, Princeton, New Jersey 08540, USA}%

\author{Chuanfei Dong}
 \email{dcfy@princeton.edu}
 \affiliation{Princeton Plasma Physics Laboratory, Princeton University, Princeton, New Jersey 08540, USA}%
\affiliation{Department of Astrophysical Sciences, Princeton University, Princeton, New Jersey 08544, USA}%
\affiliation{Department of Astronomy, Boston University, Boston, Massachusetts 02215, USA}%
 
\author{Liang Wang}
\affiliation{Department of Astrophysical Sciences, Princeton University, Princeton, New Jersey 08544, USA}%
\affiliation{Princeton Plasma Physics Laboratory, Princeton University, Princeton, New Jersey 08540, USA}%

\author{Gabor Toth}
\affiliation{Department of Climate and Space Sciences and Engineering, University of Michigan, Ann Arbor, Michigan 48109, USA}%

\date{\today}

\begin{abstract}
Collisionless magnetic reconnection typically requires kinetic treatments that are, in general, computationally expensive compared to fluid-based models. In this study, we use the magnetohydrodynamics with adaptively embedded particle-in-cell (MHD-AEPIC) model to study the interaction of two magnetic flux ropes. This innovative model embeds one or more adaptive PIC regions into a global MHD simulation domain such that the kinetic treatment is only applied in regions where kinetic physics is prominent. We compare the simulation results among three cases: 1) MHD with adaptively embedded PIC regions, 2) MHD with statically (or fixed) embedded PIC regions, and 3) a full PIC simulation. The comparison yields good agreement when analyzing their reconnection rates and magnetic island separations, as well as the ion pressure tensor elements and ion agyrotropy. In order to reach a good agreement among the three cases, large adaptive PIC regions are needed within the MHD domain, which indicates that the magnetic island coalescence problem is highly kinetic in nature where the coupling between the macro-scale MHD and micro-scale kinetic physics is important.
\end{abstract}

\maketitle

\section{\label{sec:level1} Introduction}

Magnetic reconnection is a process that occurs on the order of Alfv\'{e}nic timescales in which the magnetic topology is rearranged and the magnetic energy is converted into plasma kinetic or thermal energy \cite{Finn1977,Bondeson1983}. Because of its broad applications in solar flares \cite{Forbes1991} and corona \cite{Chen2021b,Dong2022}, coronal mass ejections (CMEs) \cite{Qiu2007}, Earth's and planetary magnetospheres \cite{Chen2008,Le2017,Phan2018,Yang2022}, and also in laboratory plasmas \cite{Yamada1994,Duan2018,Raymond2018}, reconnection has been of immense interest in recent years \cite{Ji2022}.

It has been well established that collisionless magnetic reconnection occurs in the regime beyond ideal magnetohydrodynamics, where kinetic-scale physics becomes prominent in the formation and evolution of the thin current sheets near the reconnection sites \cite{Zweibel2009,Yamada2010}. Thus, simulating collisionless magnetic reconnection typically requires kinetic approaches such as the particle-in-cell (PIC) method. In general, however, such kinetic simulations are computationally expensive and are hence unable to efficiently solve large-scale problems involving collisionless physics. In order to solve this issue with affordable computational costs, two broad approaches have been proposed for large-scale global simulations, i.e., the magnetohydrodynamics with embedded particle-in-cell (MHD-EPIC) model \cite{Daldorff2014,Toth2016,Chen2017,Chen2019JGR} and the multi-moment multi-fluid model \cite{Wang2015,Wang2018,Dong2019,Wang2020,Jarmak2020,Rulke2021}. Recently, some progress has been made to improve the fluid closure in the multi-moment multi-fluid model through machine learning \cite{Qin2022,Cheng2023}, and meanwhile, MHD-EPIC has been improved to incorporate the feature of adaptively embedded PIC regions (MHD-AEPIC) \cite{Shou2021,Chen2021,Wang2022}, which offers flexibility for the PIC code to capture the localized regions where kinetic physics is important.

It has been well demonstrated through a series of local studies \cite{Wang2015,Ng2015,Ng2017,Ng2019} that the multi-moment multi-fluid model incorporating the higher-order moments is capable of reproducing some critical aspects of the collisionless reconnection physics from fully kinetic simulations. However, no such systematic local studies have been conducted using the MHD-AEPIC model, which motivates this study to employ the MHD-AEPIC model to investigate the magnetic island coalescence problem that is highly kinetic in nature \cite{Stanier2015,Ng2015}.

In an MHD-AEPIC simulation, only part of the simulation domain, where kinetic effects are important, is simulated by the semi-implicit PIC code, Flexible Exascale Kinetic Simulator (FLEKS) \cite{Chen2021} and the rest of the domain is handled by the MHD (or Hall MHD) model, Block-Adaptive-Tree Solarwind Roe-type Upwind Scheme (BATS-R-US) \cite{Powell1999,Toth2012}. For the magnetic island coalescence problem, the embedded PIC regions are applied to simulate the regions with strong current density. As we will show later, the adaptive simulation comes in close agreement with the simulation employing the fully kinetic PIC code,  Object-oriented Simulation Rapid Implementation System (OSIRIS) \cite{Fonseca2002,Hemker2015}, when analyzing their out-of-plane current densities, reconnection rates, O-point separations, pressure tensor elements, and agyrotropy (a measure of the deviation of the pressure tensor of a species from cylindrical symmetry with respect to the direction of the local magnetic field \cite{Scudder2008}).

In Sec. II, we discuss the model setup for the magnetic island coalescence problem. In Sec. III, simulation results are presented and discussed by comparing the full PIC simulation results with the outputs from the MHD-EPIC and MHD-AEPIC models. Concluding remarks are given in Sec. IV. 

\section{Model Setup and Methods}

In this study, we set a Fadeev equilibrium \cite{Fadeev1965} as the initial condition where the initial magnetic vector potential is expressed as
\begin{eqnarray}
\textbf{A} =  B_0 \lambda \ln \big[ \epsilon \cos(x/\lambda) + \cosh(y/\lambda) \big] \hat{\textbf{z}}
\label{eq:vecpot}
\end{eqnarray}
where $B_0$ is the asymptotic magnetic field away from the x-axis, $\epsilon = 0.4$ is a measure of the island size, and the initial density distribution is written as
\begin{eqnarray}
n = \frac{n_0 (1 - \epsilon^2)}{\big[ \epsilon \cos(x/\lambda) + \cosh(y/\lambda) \big]^2} + n_b
\label{eq:eqden}.
\end{eqnarray}
where $n_b = 0.2n_0$ is the background number density. We define the global Alfv\'{e}n time as $t_A = L_x/v_A$, where $L_x = 4\pi \lambda$ is the length of the simulation box, $\lambda = 5d_{i0}$ is the half-width of the current sheet (we also run additional simulations with $\lambda = 10d_{i0}$; see Fig.~\ref{fig:lambda10_rate}), $d_{i0} = (m_i/\mu_0 n_0 e^2)^{1/2}$ is the ion inertial length, and $v_A = B_0/(\mu_0 n_0 m_i)^{1/2}$ is the Alfv\'{e}n speed. The width of the box is set as $L_y = 2\pi \lambda$. In the full PIC simulation with OSIRIS, we specify the ion and electron out-of-plane current densities as the following relationship $J_{zi0}/J_{ze0} = T_{i0}/T_{e0}$, where $T_{i0}$ and $T_{e0}$ are uniform. In the MHD-AEPIC simulations, the electron and ion bulk velocities are initialized from the MHD current and momentum \cite{Daldorff2014}. In the following simulations, we use the same configuration described in \citet{Stanier2015} and \citet{Ng2015}, setting $T_{i0} = T_{e0}$, $m_i = 25m_e$ (so $d_i =  5 d_e$), $t_{max} = 2.5 t_A$, and $n_0 k_B (T_{i0} + T_{e0}) = B_0^2/2\mu_0$ (so $p = p_{mag}$). The relationship between the electron plasma and cyclotron frequencies is given by $\omega_{pe} = 2\Omega_{ce}$ where $\omega_{pe} = (n_0 e^2/m_e \epsilon_0)^{1/2}$ and $\Omega_{ce} = |e|B_0/m_e$. The initial perturbation follows the same form as in \citet{Daughton2009_1}, with
\begin{subequations}
\label{eq:idealMHDpert}
\begin{eqnarray}
\delta B_x = \delta B_0 \cos \bigg( \frac{2 \pi x}{L_x} \bigg) \sin \bigg( \frac{\pi y}{L_y} \bigg)
\label{magpert1}
\end{eqnarray}
\begin{eqnarray}
\delta B_y = -\delta B_0 \sin \bigg( \frac{2 \pi x}{L_x} \bigg) \cos \bigg( \frac{\pi y}{L_y} \bigg)
\label{magpert2}
\end{eqnarray}
\end{subequations}
where we set the perturbation amplitude as $\delta B_0 = 0.1B_0$.

To simulate the merging of two magnetic islands, we first perform two simulations using the MHD-AEPIC model, one with a large static, non-adaptive (or fixed) PIC region and the other with adaptive PIC regions. We also run a full PIC simulation using OSIRIS to validate and compare with the MHD-AEPIC simulations. In all simulations, we have used periodic boundaries in the horizontal ($x$) direction, and conducting electromagnetic boundaries and reflecting particle boundaries in the vertical ($y$) direction.

The OSIRIS simulation domain consists of 2000$\times$1000 cells and each cell has 256 particles per species. In the MHD-AEPIC simulations, the MHD grid consists of 2048$\times$1024 cells. The effective PIC grid resolution and the initial particle number per cell are the same as the OSIRIS simulation, and the Gauss's Law-satisfying Energy Conserving Semi-Implicit Method (GL-ECSIM) \cite{Chen2019} is applied. In general, the MHD and PIC grid size does not have to be the same \cite{Chen2021} (see Appendix A). In the magnetic island coalescence simulation, the two magnetic islands are associated with strong current density, which suggests a notable separation between the electron and ion velocities, and kinetic effects may play an important role when two islands move toward each other. Outside the islands, the current is generally weak, and neither the magnetic field nor plasma quantities show a strong gradient, so an MHD model is sufficient to describe this region. In the MHD-AEPIC run with adaptive PIC regions, the PIC regions only cover the areas of $J\Delta x/B > 0.01$, where the selection criterion $J\Delta x/B$ is dimensionless. The threshold 0.01 is carefully chosen based on numerical experiments such that the islands are well covered by the PIC region. The selection criterion is calculated from the MHD results. We usually run a pure MHD simulation first to estimate a proper threshold that would determine the desired PIC regions, then we run short MHD-AEPIC simulations with a coarse PIC grid to optimize the threshold. The small threshold for the island coalescence problem is due to the fact that it is highly kinetic in nature where the coupling between the macro-scale MHD and micro-scale kinetic physics is important \cite{Stanier2015,Ng2019}. In general, the PIC region of an MHD-AEPIC simulation is selected based on the nature of a problem. For comparison, we also perform a simulation using the MHD-EPIC model with a relatively large static or fixed PIC region, which covers the majority of the entire simulation domain. 

In MHD-AEPIC and MHD-EPIC simulations, when Hall MHD is used (which is the case for the current study), the Alfven and Whistler modes can travel through the MHD-PIC boundaries smoothly \cite{Daldorff2014}. However, if there are kinetic modes reaching the MHD-PIC boundaries, they may be artificially reflected since they cannot propagate into the MHD regions. For such applications, we should increase the PIC regions so that the kinetic modes would have been damped near the PIC regions. The MHD-AEPIC model is initially designed for global magnetosphere simulations with a relatively small kinetic region. If the entire simulation domain is dominated by kinetic physics, a full PIC or hybrid-PIC \cite{Dong2021,Winske2022} code is a better choice. 

We note that the boundary conditions, with the exception of periodic boundaries, introduce numerical perturbations inevitably for all kinetic codes, which is a natural consequence of numerical discretization. The perturbations, however, would be reduced if the boundaries are far away from the kinetic regions. We, therefore, usually choose PIC regions where the solution is smooth at the MHD-PIC interface, such as the simulations presented in this paper.

\section{Simulation Results}

\begin{figure}
\includegraphics[width=0.5\textwidth]{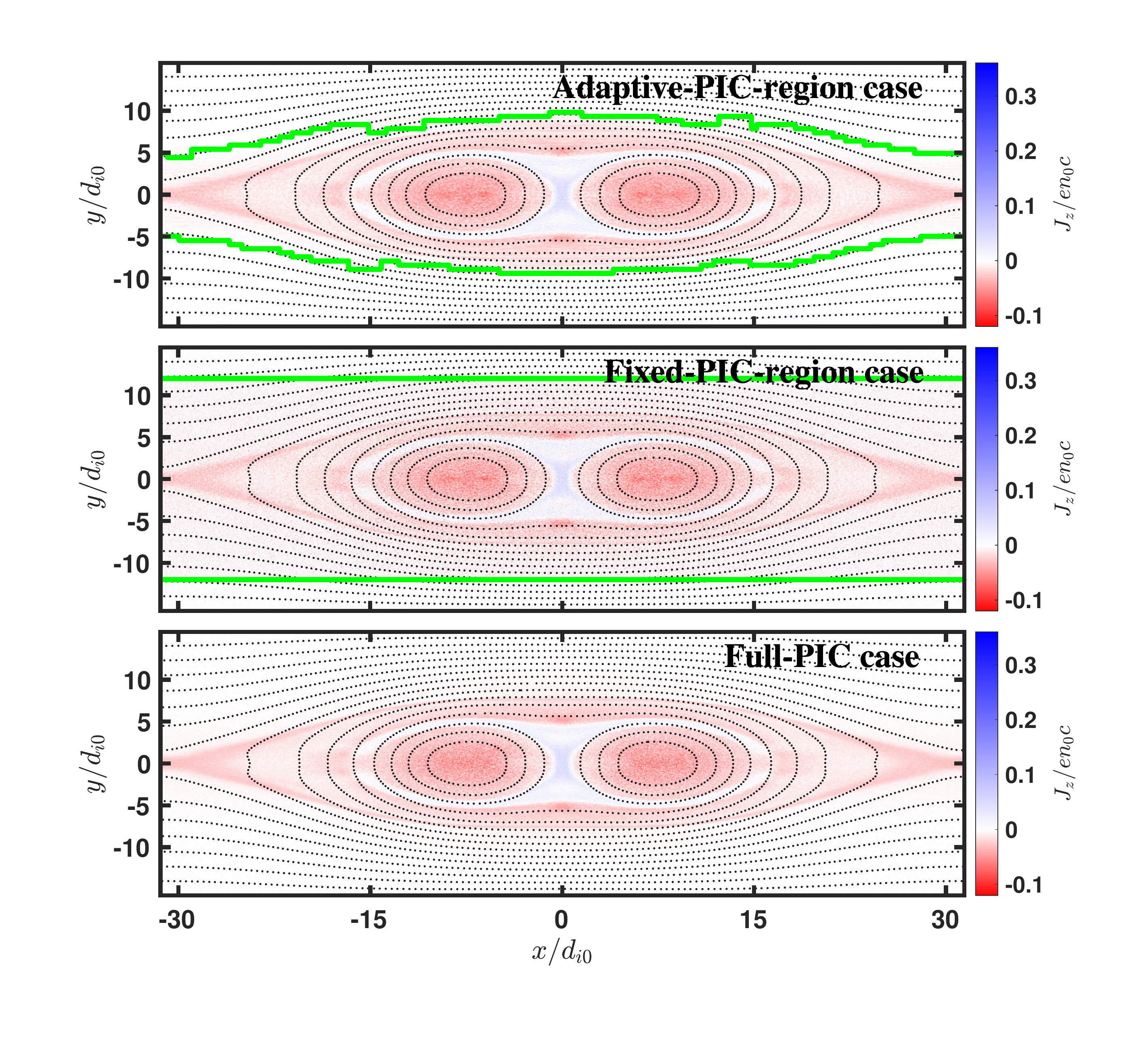}
\caption{\label{fig:Jz-contour} Out-of-plane current density at $t = 0.9 t_A$ for the adaptive (top) and fixed (middle) PIC region runs and the full-PIC run (bottom) with $\lambda=5 d_{i0}$, overlaid with the magnetic vector potential contours (black dotted curves) and with the boundary between the MHD and PIC regions (green solid curves).}
\end{figure}

\begin{figure}
\includegraphics[width=0.5\textwidth]{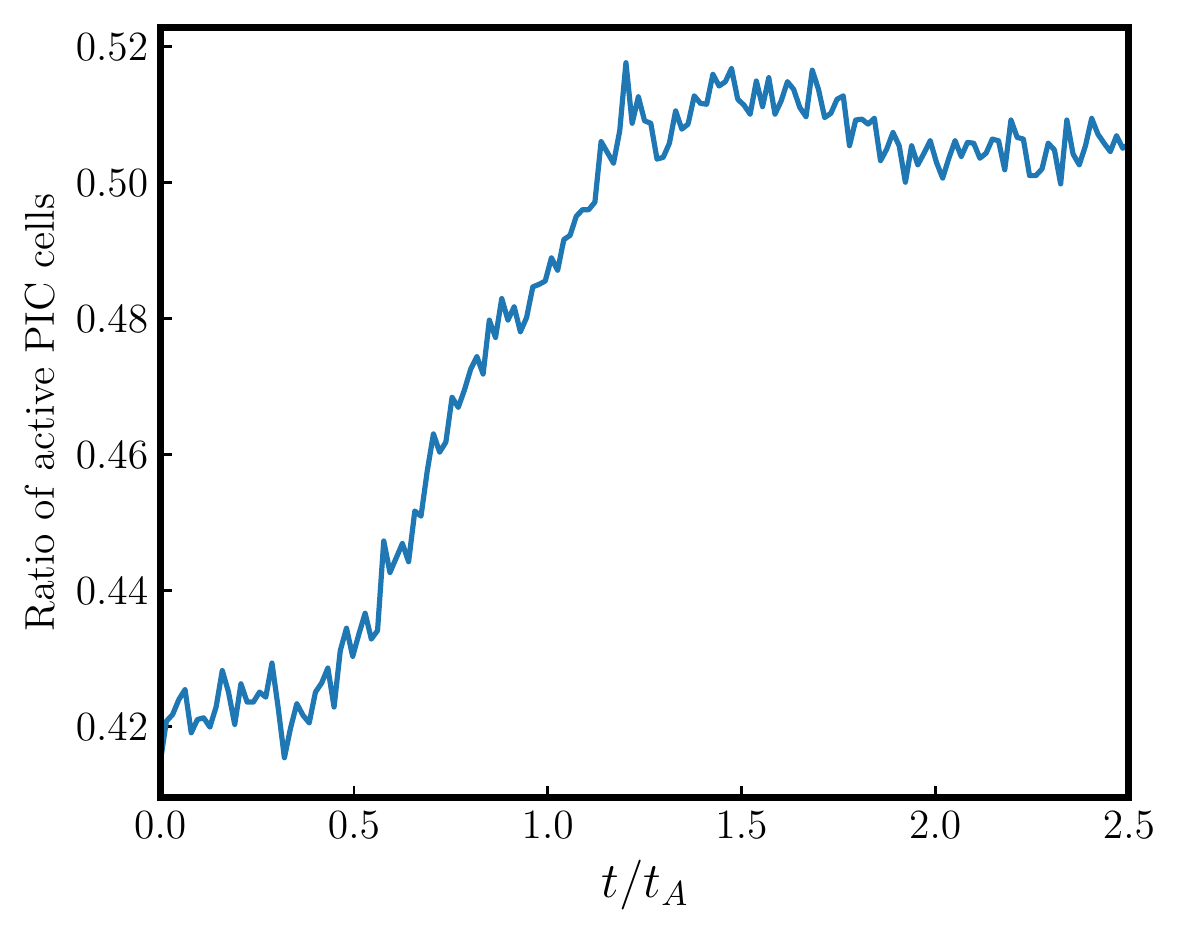}
\caption{\label{fig:pic_cell_ratio} Ratio of the active PIC regions over the entire simulation domain for the MHD-AEPIC simulation. The regions that are covered by the semi-implicit PIC code increase from $40\%$ to about $50\%$ during the simulation.
}
\end{figure}

Fig.~\ref{fig:Jz-contour} shows 2D plots of the out-of-plane current density at $t=0.9t_A$ from the three different simulations, where the $x$ and $y$ coordinates have been normalized over $d_{i0}$ and the current density over $|e| n_0 c$. In each panel of Fig.~\ref{fig:Jz-contour}, the current sheet formation, which cannot be correctly captured in a Hall MHD model for this problem \cite{Stanier2015}, are indicated as regions of positive out-of-plane current densities near the origin ($x, y = 0$). Also illustrated in Fig.~\ref{fig:Jz-contour} are the boundaries (green curves) between the MHD and PIC regions in both the adaptive-PIC-region and fixed-PIC-region runs. 

For the adaptive-PIC-region run, the variation of the active PIC cell number inside the entire simulation domain over time is shown in Fig.~\ref{fig:pic_cell_ratio}. Initially, about 40\% of the simulation domain is run by PIC code, then the ratio gradually increases to about 50\%. In this simulation, the PIC regions are defined as the areas with large current density values and they do not vary significantly during the simulation. For other applications, such as modeling the Earth magnetotail \cite{Wang2022}, the active PIC regions may change dramatically. We note that the smallest granularity to turn on or off PIC cells is a patch with two cells in each direction \cite{Chen2021}. Since the MHD-AEPIC model adopts a semi-implicit PIC code, which requires an iterative solver to update the electric field \cite{Chen2019}, it requires more computations per step. The computational efficiency also depends on the implementation and problem properties. For the simulations presented in Fig.~\ref{fig:Jz-contour}, their computational cost (CPU hours) ratio is about $3:8:1$ among the adaptive-PIC region case, the fixed-PIC region case, and the full-PIC case. Given that FLEKS is a semi-implicit PIC code and the gird resolution between the PIC and MHD regions can be different, one can reduce the grid resolution of the PIC cells inside the MHD-AEPIC domain, which can lower the computational cost of the MHD-AEPIC run and thus the MHD-AEPIC run becomes computationally cheaper than the full-PIC case (see Appendix A). It is noteworthy that in three-dimensional (3-D) cases, MHD-AEPIC can be computationally even more efficient than a full PIC code. 

\begin{figure}
\includegraphics[width=0.5\textwidth]{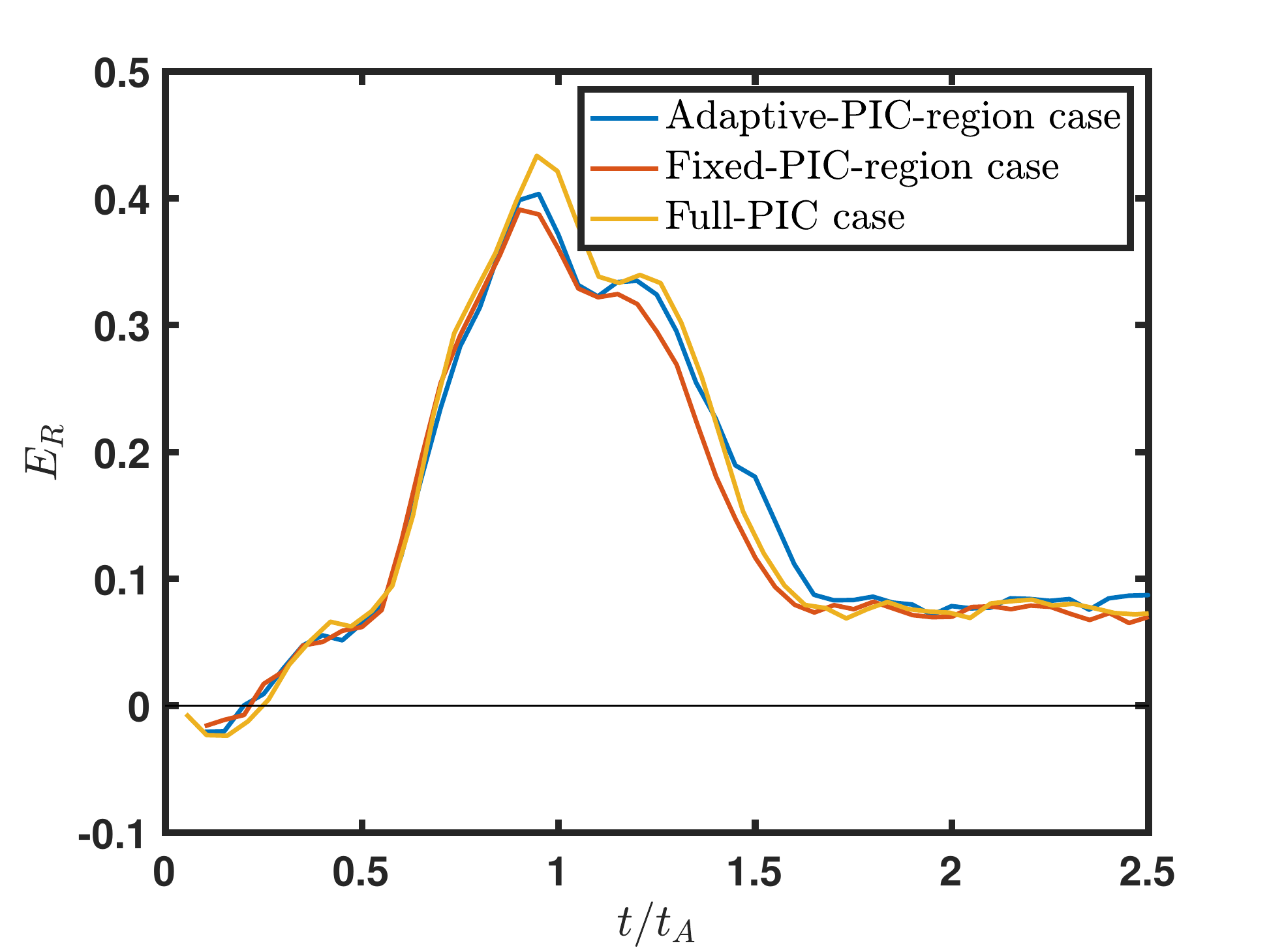}
\caption{\label{fig:ER} Reconnection rate as a function of time for the adaptive and fixed PIC region runs and the full-PIC run with $\lambda=5 d_{i0}$.}
\end{figure}

\begin{figure}
\includegraphics[width=0.5\textwidth]{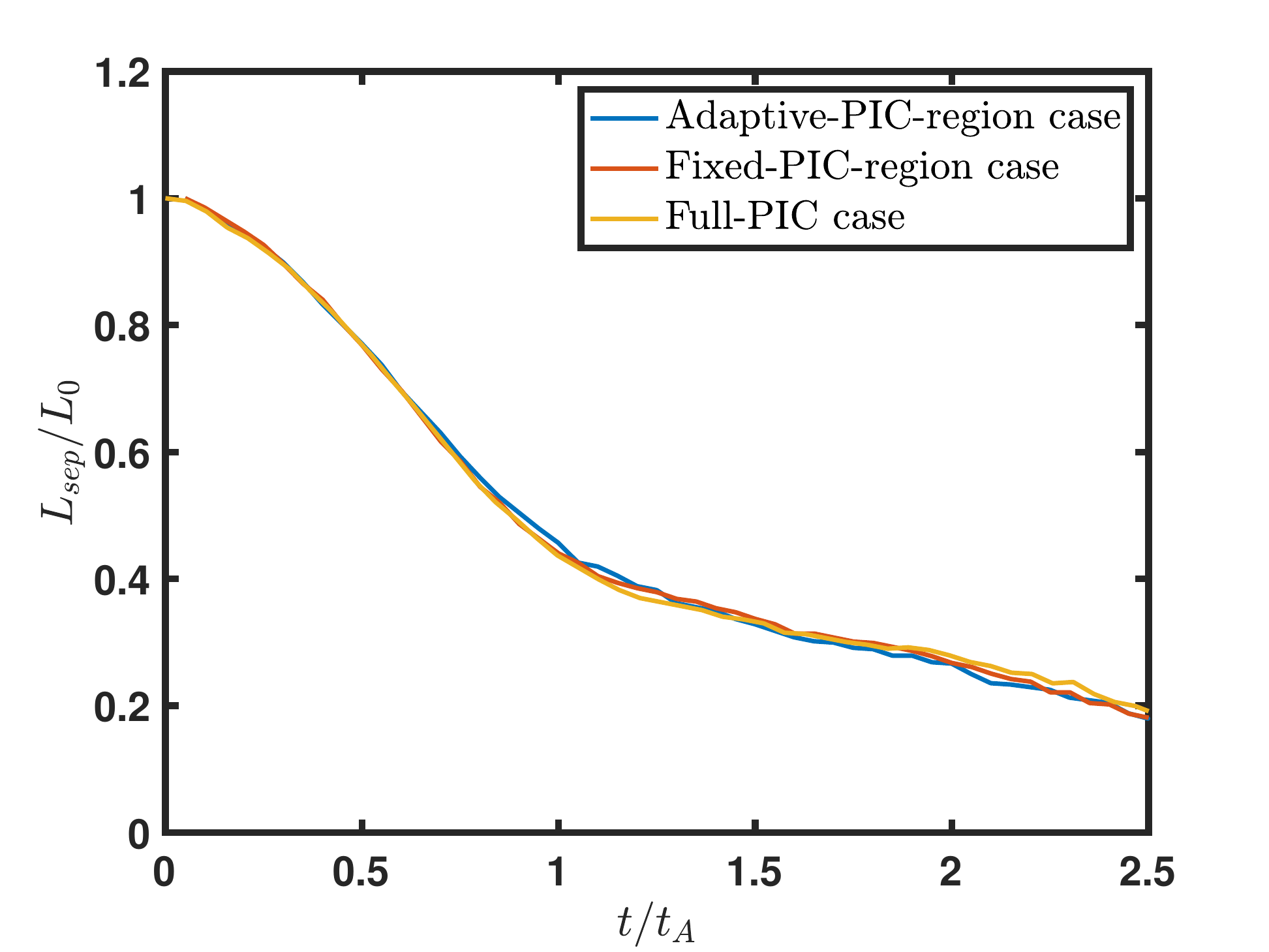}
\caption{\label{fig:O-sep} O-point separation of the coalescing magnetic islands as a function of time for the adaptive and fixed PIC region runs and the full-PIC run with $\lambda=5 d_{i0}$.}
\end{figure}

An inspection of Fig.~\ref{fig:Jz-contour} reveals that the simulation results (such as the current density and the magnetic flux) from three different runs present nearly identical features and the system evolves at the same rate, indicating the reconnection rates, $E_R$, from the three simulations are very similar. In order to verify this idea, we explicitly compare the reconnection rates. We normalize the reconnection rate to the maximum initial magnetic field between the islands such that \cite{Stanier2015}
\begin{eqnarray}
E_R = \frac{1}{v_{Am}B_m} \frac{\partial}{\partial t} \big( A_{zX} - A_{zO} \big)
\label{eq:three}
\end{eqnarray}
where $v_{Am} = B_m/(\mu_0 n_0 m_i)^{1/2}$, $A_{zX}$ is the out-of-plane magnetic vector potential, evaluated at the $X$-point and $A_{zO}$ is evaluated at the $O$-point. The time evolution of the reconnection rate for the three simulations are plotted in Fig.~\ref{fig:ER}, where a maximum reconnection rate occurs around $t = 0.9 t_A$ for all the cases. After the reconnection rate reaches a maximum, it decreases with time before reaching a constant value at approximately $t = 1.6 t_A$. We note from Fig.~\ref{fig:ER} that the case with adaptive PIC regions closely captures the behavior of $E_R$ for this problem, when compared to the full-PIC simulation.

Since the O-point separation between the magnetic islands is largely dependent on reconnection rates, we demonstrate the O-point separation as a function of time in Fig.~\ref{fig:O-sep}. In this plot, we have normalized the O-point separation by the initial separation between the islands $L_0$. It is evident in Fig.~\ref{fig:O-sep} that for all three cases the island separation decreases at an increasing rate before $t = 0.9 t_A$, at which point the coalescence of the islands slows down, reaching an approximately constant rate at $t = 1.6 t_A$. Both Figs.~\ref{fig:ER} and \ref{fig:O-sep} show excellent agreement among the three simulation approaches, validating the use of the MHD-AEPIC model.

\begin{figure*}
\includegraphics[width=\textwidth]{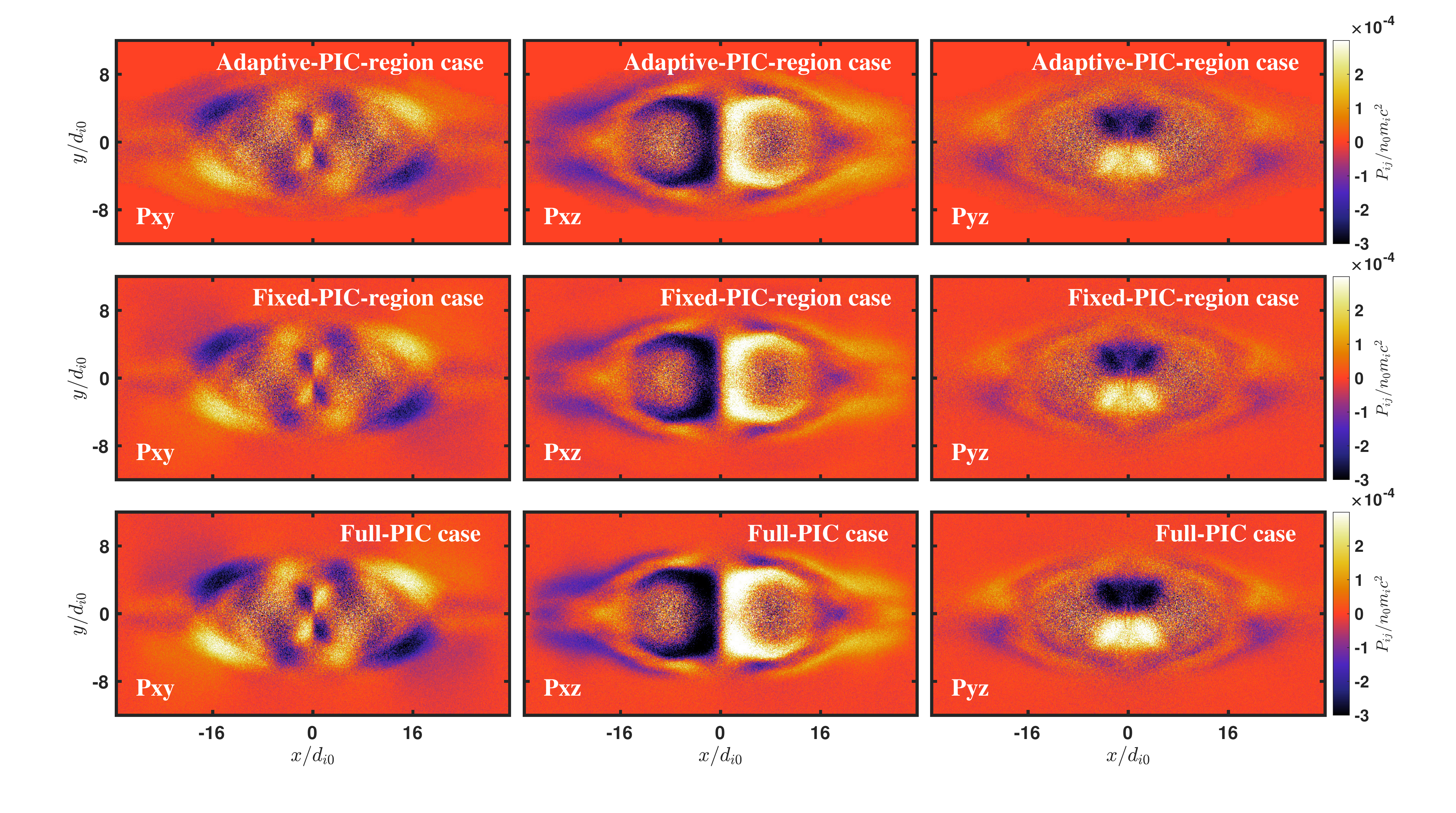}
\caption{\label{fig:fig4} Comparison of the off-diagonal ion pressure tensor elements at $t = 0.9 t_A$ for the adaptive (top row) and fixed (middle row) PIC region runs and the full-PIC run (bottom row) with $\lambda=5 d_{i0}$. The left column plots $P_{xy}$, the middle column shows $P_{xz}$, and $P_{yz}$ is illustrated in the right column.}
\end{figure*}

At this point, we are interested in analyzing the effects of ion kinetics, as its importance has been demonstrated previously \cite{Stanier2015}. \citet{Stanier2015} also pointed out that electron kinetics were not crucial for the island coalescence problem. To do this, we write the z-component of the normalized ion Ohm's law as 
\begin{eqnarray}
E_z^\prime &=& \frac{d_i}{n} \bigg[ \frac{\partial}{\partial t} (nv_{iz}) + \nabla \cdot (n\textbf{v}_i v_{iz}) \bigg] + \frac{d_i}{n} \nabla \cdot \big( \bar{\bar{\textbf{P}}}_i \cdot \textbf{z} \big)
\label{eq:four}
\end{eqnarray}
where $E_z^\prime = (\textbf{E} + \textbf{v}_i \times \textbf{B}) \cdot \textbf{z}$ is the z-component of the nonideal electric field, $\bar{\bar{\textbf{P}}}_i$ is the ion pressure tensor, and the resistivity is neglected for collisionless reconnection. Near the X-point of the reconnection site, where the magnetic field is small, pressure tensor effects become important. Specifically, Eq.(\ref{eq:four}) demonstrates that the non-ideal electric field is heavily influenced by the off-diagonal terms of the pressure tensor. Since it has been shown in \citet{Stanier2015} that ion kinetic effects are of importance in this specific reconnection setup, we focus our analysis on the ion pressure tensor. Fig.~\ref{fig:fig4} compares the off-diagonal ion pressure tensor elements from the three cases when the reconnection rate reaches its peak, once again showing spectacular agreement among the cases. Fig.~\ref{fig:fig4} also demonstrates that these two islands should be covered by the PIC code, since the off-diagonal pressure tensor can not be described by the isotropic MHD used here. In general, if the ion pressure is anisotropic, but the off-diagonal terms are relatively small in the local magnetic field coordinate system, MHD-AEPIC supports coupling with an anisotropic MHD model \cite{Daldorff2014}, and the region that requires PIC can be reduced. However, if the off-diagonal terms are prominent in the local magnetic field coordinates, which is the case here (see Appendix B), the PIC regions need to cover those areas, consistent with the previous study using a similar approach \cite{Makwana2018}.

\begin{figure}
\includegraphics[width=0.5\textwidth]{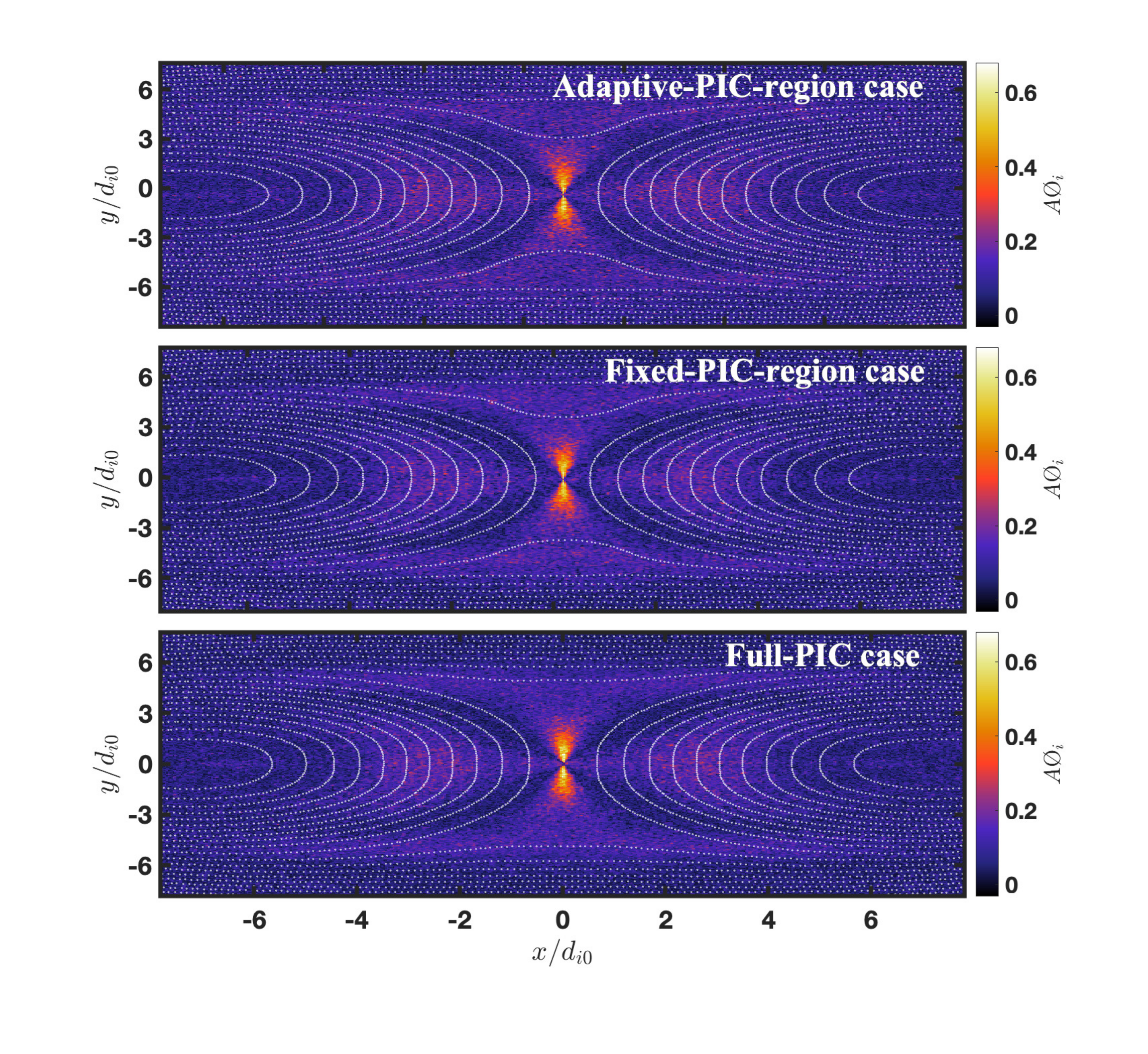}
\caption{\label{fig:agyrotropy} Ion agyrotropy at $t = 0.9 t_A$ for the adaptive (top) and fixed (middle) PIC region runs and the full-PIC run (bottom) with $\lambda=5 d_{i0}$, overlaid with the magnetic vector potential contours (in white).}
\end{figure}

In addition to the off-diagonal ion pressure tensor terms, we also compare the ion agyrotropies of the three simulations, employing the same formula for agyrotropy as in \citet{Scudder2008}. We note that agyrotropy is a measure useful for characterizing the ion (or electron) diffusion region in collisionless magnetic reconnection. We once again focus only on ion agyrotropy as ion kinetics are required to capture the correct reconnection rates and describe the global behavior of the system for the island coalescence problem \cite{Stanier2015}. A comparison of the agyrotropy from the three different simulations is shown in Fig.~\ref{fig:agyrotropy}, indicating excellent agreement among different cases, especially near the current sheet formation where ion agyrotropy is of particular significance (indicated by the lighter colors inside the ion diffusion region). 

\begin{figure}
\includegraphics[width=0.5\textwidth]{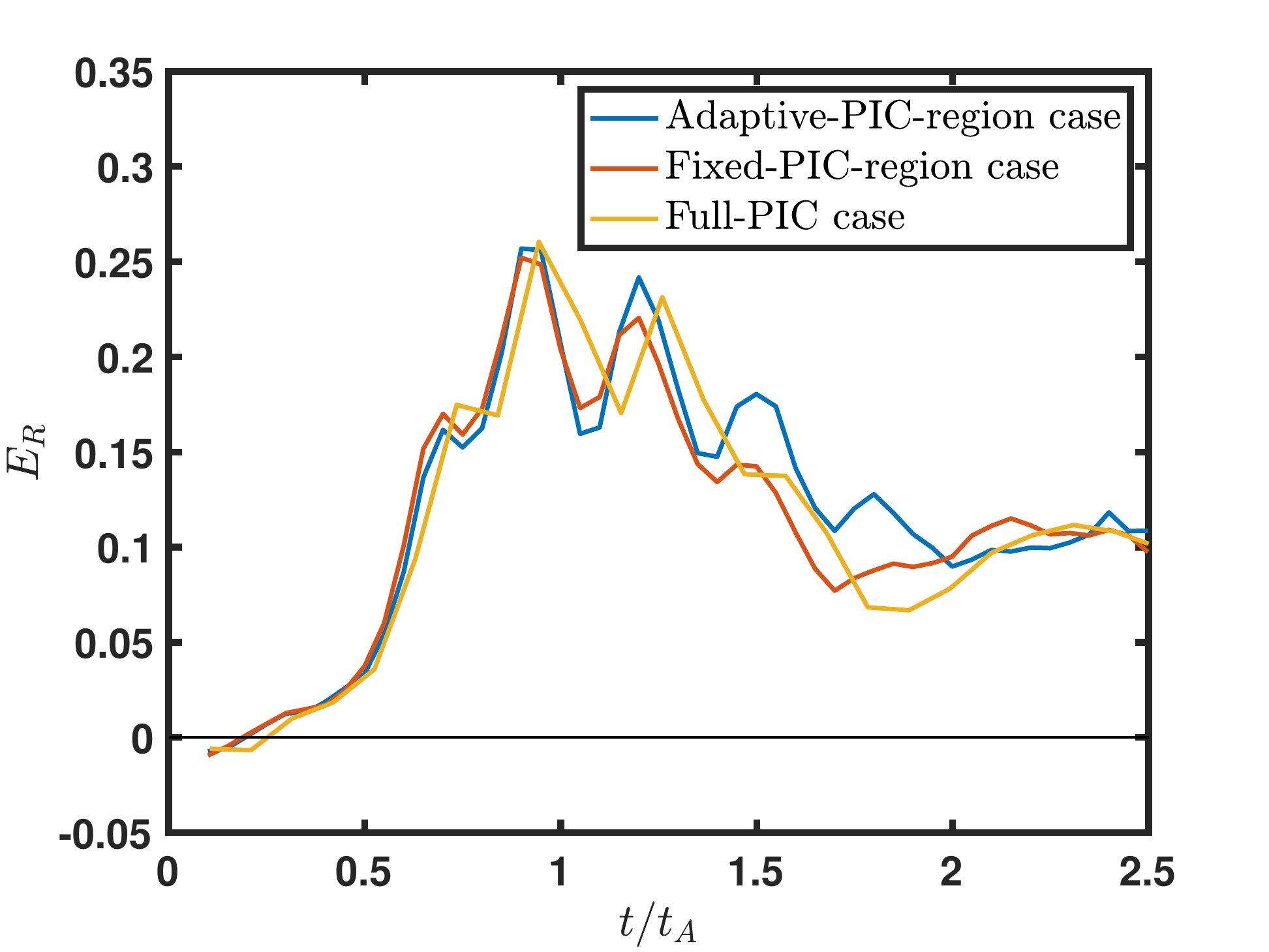}
\caption{\label{fig:lambda10_rate} Reconnection rate as a function of time for the adaptive and fixed PIC region runs and the full-PIC run with $\lambda=10 d_{i0}$.}
\end{figure}

In Fig.~\ref{fig:lambda10_rate}, we also present the reconnection rates for simulations with $\lambda = 10 d_{i0}$ to demonstrate that reconnection in large systems becomes slower. The islands are unable to coalesce on the first approach due to the slower reconnection rates, and so bounce off each other \cite{Karimabadi2011_1,Stanier2015}. All simulations reach nearly the same maximum reconnection rate and show two peaks around $t=0.9t_A$ and $t=1.2t_A$, respectively. Overall, all simulations show similar reconnection processes.

\section{Conclusion}

It has been shown that simulating collisionless magnetic reconnection requires kinetic approaches such as the PIC method. However, these simulation codes are generally computationally expensive and hence make it difficult to efficiently model large-scale global systems. In this study, we have presented an alternative method by utilizing the MHD-AEPIC model that uses the PIC treatment in regions where the current sheet formation is prominent and uses the computationally cheap MHD model in the rest of the simulation domain. We have shown that the case with adaptive PIC regions comes in close agreement with the full PIC simulation when analyzing reconnection rates and O-point separations as well as the ion pressure tensor elements and ion agyrotropy. These results demonstrate that the MHD-AEPIC model may accurately and efficiently simulate large-scale systems that involve collisionless reconnection physics. 

It should be noted that the magnetic island coalescence problem that we studied here is highly kinetic in nature where the coupling between the MHD and kinetic scales is important \cite{Daughton2009_2,Stanier2015,Ng2019}, so relatively large PIC regions are needed within the MHD domain. For large-scale simulations such as the case of the Earth's magnetosphere, \citet{Chen2021} applied MHD-AEPIC to study the magnetopause reconnection while treating the magnetotail reconnection using Hall MHD, such that they drastically saved computational costs compared with the full PIC magnetosphere simulations.

We also note that in magnetic confinement fusion, near the X-point where the magnetic field is weak, gyrokinetic approximations begin to break down and it becomes necessary to include full kinetic physics at these locations. The MHD-AEPIC model may be useful for simulating FLARE \cite{Ji2020} and NSTX-U \cite{Ebrahimi2016} at Princeton Plasma Physics Laboratory, MAST at the UK \cite{Tanabe2017} as well as TREX at Wisconsin Plasma Physics Laboratory \cite{Olson2021}.

\begin{figure}
\includegraphics[width=0.5\textwidth]{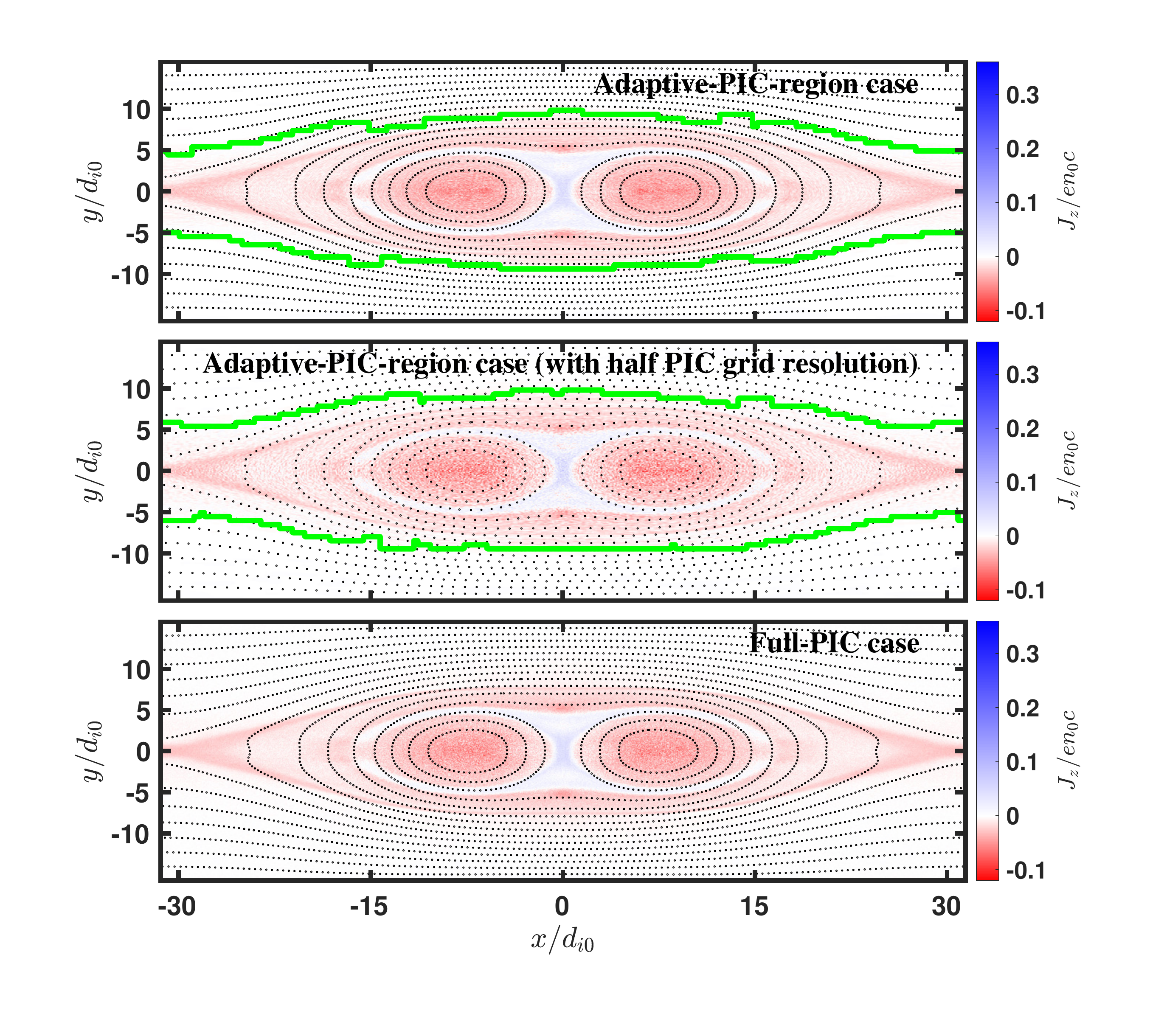}
\caption{\label{fig:lambda5_halfres_boundaries} Out-of-plane current density at $t = 0.9 t_A$ for the adaptive PIC region run (top), the adaptive PIC region run with half PIC grid resolution (middle), and the full-PIC run (bottom) with $\lambda=5 d_{i0}$, overlaid with the magnetic vector potential contours (black dotted curves) and with the boundary between the MHD and PIC regions (green solid curves).}
\end{figure}

\begin{figure}[b]
\includegraphics[width=0.5\textwidth]{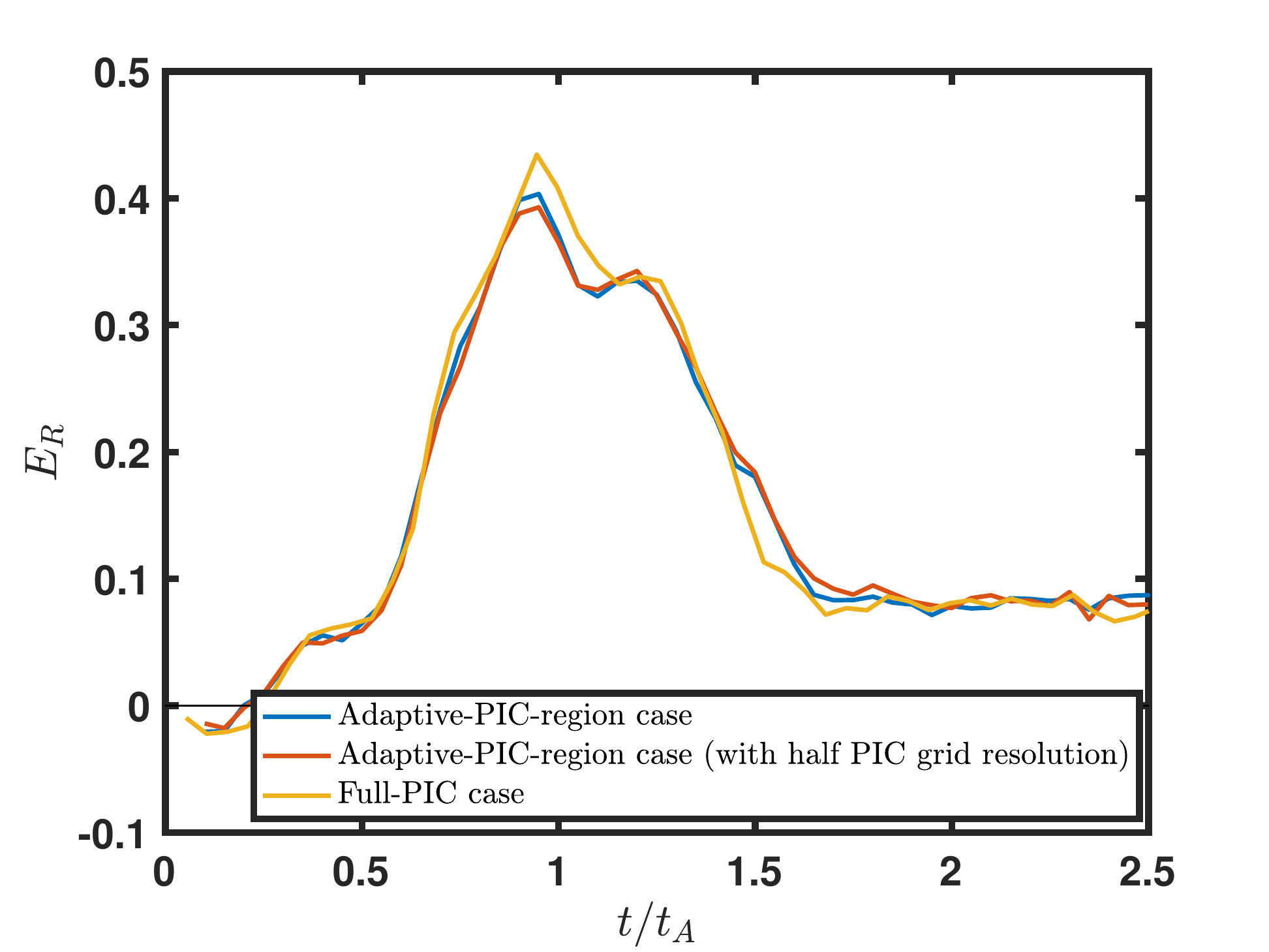}
\caption{\label{fig:lambda5_halfres_rate} Reconnection rate as a function of time for the three cases shown in Fig.\ref{fig:lambda5_halfres_boundaries}.}
\end{figure}

\begin{figure*}
\includegraphics[width=\textwidth]{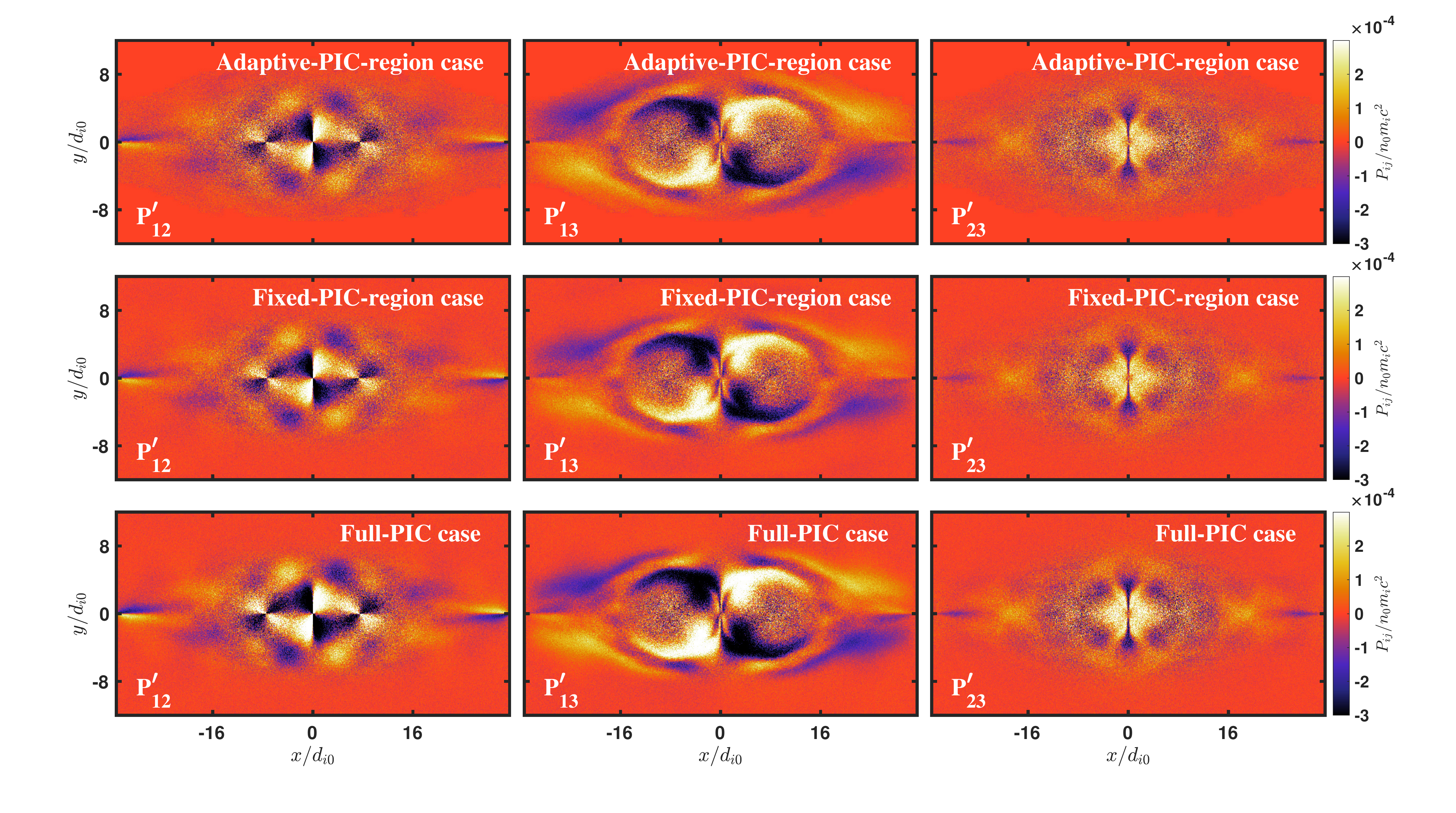}
\caption{\label{fig:fieldaligned} Comparison of the off-diagonal ion pressure tensor elements at $t = 0.9 t_A$ for the adaptive (top row) and fixed (middle row) PIC region runs and the full-PIC run (bottom row) in field-aligned coordinates with $\lambda = 5 d_{i0}$. The left column plots $P_{12}^\prime$, the middle column shows $P_{13}^\prime$, and $P_{23}^\prime$ is illustrated in the right column.}
\end{figure*}

\begin{acknowledgments}

This work was made possible by support from the Department of Energy for the Summer Undergraduate Laboratory Internship (SULI) program. This work was supported by the U.S. Department of Energy under contract number DE-AC02-09CH11466 (through LDRD) and the DOE grant DE-SC0021205, the NASA grants 80NSSC21K1326 and 80NSSC22K0323, and the NSF grant AGS-2149787. Resources supporting this work were provided by the NASA High-End Computing (HEC) Program through the NASA Advanced Supercomputing (NAS) Division at Ames Research Center. The authors acknowledge the OSIRIS Consortium, consisting of UCLA and IST (Portugal) for the use of the OSIRIS 4.0 framework.

\end{acknowledgments}

\section*{Data Availability}
The data that support the findings of this study are available from the corresponding author upon reasonable request.

\appendix

\section{MHD-AEPIC Simulations with Different PIC Grid Resolutions}

Fig.~\ref{fig:lambda5_halfres_boundaries} compares the out-of-plane current densities at $t = 0.9 t_A$ among the adaptive-PIC-region case, the adaptive-PIC-region case (with half PIC grid resolution), and the full-PIC simulation. The MHD grid resolution between the two adaptive PIC cases is the same, which demonstrates the flexibility and capability of the coupling algorithm between the MHD model and the semi-implicit PIC code. As shown in Figs.~\ref{fig:lambda5_halfres_boundaries} \&~\ref{fig:lambda5_halfres_rate}, all the simulations show very similar results. In general, a semi-implicit PIC code allows relaxed stability constraints, therefore, it can use a coarser grid resolution than an explicit PIC code.

\section{Off-Diagonal Pressure Tensor Elements in Field-Aligned Coordinates}

Fig.~\ref{fig:fieldaligned} depicts the off-diagonal ion pressure tensor elements in field-aligned coordinates (or local magnetic field coordinates) for the three cases at $t = 0.9 t_A$, showing that the off-diagonal terms are prominent in the local magnetic field coordinates. The rotated pressure tensor elements $P^\prime_{mn}$ in field-aligned coordinates are defined as follows,

\begin{eqnarray}
P^\prime_{mn} = (R^TPR)_{mn} = R_{im} P_{ij} R_{jn} 
\end{eqnarray}
where $P_{ij}$ are the non-rotated ion pressure tensor elements, the rotation matrix is

\begin{eqnarray}
\overleftrightarrow{\textbf{R}} = 
\begin{pmatrix}
b_x & -b_y/\sqrt{b_x^2 + b_y^2} & -b_xb_z/\sqrt{(b_x^2 + b_y^2)^2 + b_x^2b_z^2 + b_y^2b_z^2} \\
b_y & b_x/\sqrt{b_x^2 + b_y^2} & -b_yb_z/\sqrt{(b_x^2 + b_y^2)^2 + b_x^2b_z^2 + b_y^2b_z^2} \\ \nonumber
b_z & 0 & (b_x^2 + b_y^2)/\sqrt{(b_x^2 + b_y^2)^2 + b_x^2b_z^2 + b_y^2b_z^2} \\ 
\end{pmatrix}
\end{eqnarray}
\begin{eqnarray}
\end{eqnarray}
and $\textbf{b}$ is the unit vector along the local magnetic field direction.

\section*{REFERENCES}
\nocite{*}


\begin{thebibliography}{52}%
\makeatletter
\providecommand \@ifxundefined [1]{%
 \@ifx{#1\undefined}
}%
\providecommand \@ifnum [1]{%
 \ifnum #1\expandafter \@firstoftwo
 \else \expandafter \@secondoftwo
 \fi
}%
\providecommand \@ifx [1]{%
 \ifx #1\expandafter \@firstoftwo
 \else \expandafter \@secondoftwo
 \fi
}%
\providecommand \natexlab [1]{#1}%
\providecommand \enquote  [1]{``#1''}%
\providecommand \bibnamefont  [1]{#1}%
\providecommand \bibfnamefont [1]{#1}%
\providecommand \citenamefont [1]{#1}%
\providecommand \href@noop [0]{\@secondoftwo}%
\providecommand \href [0]{\begingroup \@sanitize@url \@href}%
\providecommand \@href[1]{\@@startlink{#1}\@@href}%
\providecommand \@@href[1]{\endgroup#1\@@endlink}%
\providecommand \@sanitize@url [0]{\catcode `\\12\catcode `\$12\catcode
  `\&12\catcode `\#12\catcode `\^12\catcode `\_12\catcode `\%12\relax}%
\providecommand \@@startlink[1]{}%
\providecommand \@@endlink[0]{}%
\providecommand \url  [0]{\begingroup\@sanitize@url \@url }%
\providecommand \@url [1]{\endgroup\@href {#1}{\urlprefix }}%
\providecommand \urlprefix  [0]{URL }%
\providecommand \Eprint [0]{\href }%
\providecommand \doibase [0]{http://dx.doi.org/}%
\providecommand \selectlanguage [0]{\@gobble}%
\providecommand \bibinfo  [0]{\@secondoftwo}%
\providecommand \bibfield  [0]{\@secondoftwo}%
\providecommand \translation [1]{[#1]}%
\providecommand \BibitemOpen [0]{}%
\providecommand \bibitemStop [0]{}%
\providecommand \bibitemNoStop [0]{.\EOS\space}%
\providecommand \EOS [0]{\spacefactor3000\relax}%
\providecommand \BibitemShut  [1]{\csname bibitem#1\endcsname}%
\let\auto@bib@innerbib\@empty
\bibitem [{\citenamefont {Finn}\ and\ \citenamefont {Kaw}(1977)}]{Finn1977}%
  \BibitemOpen
  \bibfield  {author} {\bibinfo {author} {\bibfnamefont {J.~M.}\ \bibnamefont
  {Finn}}\ and\ \bibinfo {author} {\bibfnamefont {P.~K.}\ \bibnamefont {Kaw}},\
  }\bibfield  {title} {\enquote {\bibinfo {title} {Coalescence instability of
  magnetic islands},}\ }\href@noop {} {\bibfield  {journal} {\bibinfo
  {journal} {The Physics of Fluids}\ }\textbf {\bibinfo {volume} {20}},\
  \bibinfo {pages} {72--78} (\bibinfo {year} {1977})},\ \bibinfo {note}
  {{DOI:10.1063/1.861709}}\BibitemShut {NoStop}%
\bibitem [{\citenamefont {Bondeson}(1983)}]{Bondeson1983}%
  \BibitemOpen
  \bibfield  {author} {\bibinfo {author} {\bibfnamefont {A.}~\bibnamefont
  {Bondeson}},\ }\bibfield  {title} {\enquote {\bibinfo {title} {Linear
  analysis of the coalescence instability},}\ }\href@noop {} {\bibfield
  {journal} {\bibinfo  {journal} {The Physics of Fluids}\ }\textbf {\bibinfo
  {volume} {26}},\ \bibinfo {pages} {1275--1278} (\bibinfo {year} {1983})},\
  \bibinfo {note} {{DOI:10.1063/1.864287}}\BibitemShut {NoStop}%
\bibitem [{\citenamefont {Forbes}(1991)}]{Forbes1991}%
  \BibitemOpen
  \bibfield  {author} {\bibinfo {author} {\bibfnamefont {T.~G.}\ \bibnamefont
  {Forbes}},\ }\bibfield  {title} {\enquote {\bibinfo {title} {Magnetic
  reconnection in solar flares},}\ }\href@noop {} {\bibfield  {journal}
  {\bibinfo  {journal} {Geophysical $\&$ Astrophysical Fluid Dynamics}\
  }\textbf {\bibinfo {volume} {62}},\ \bibinfo {pages} {15--36} (\bibinfo
  {year} {1991})}\BibitemShut {NoStop}%
\bibitem [{\citenamefont {{Chen}}\ \emph {et~al.}(2021)\citenamefont {{Chen}},
  \citenamefont {{Przybylski}}, \citenamefont {{Peter}}, \citenamefont
  {{Tian}}, \citenamefont {{Auch{\`e}re}},\ and\ \citenamefont
  {{Berghmans}}}]{Chen2021b}%
  \BibitemOpen
  \bibfield  {author} {\bibinfo {author} {\bibfnamefont {Y.}~\bibnamefont
  {{Chen}}}, \bibinfo {author} {\bibfnamefont {D.}~\bibnamefont
  {{Przybylski}}}, \bibinfo {author} {\bibfnamefont {H.}~\bibnamefont
  {{Peter}}}, \bibinfo {author} {\bibfnamefont {H.}~\bibnamefont {{Tian}}},
  \bibinfo {author} {\bibfnamefont {F.}~\bibnamefont {{Auch{\`e}re}}}, \ and\
  \bibinfo {author} {\bibfnamefont {D.}~\bibnamefont {{Berghmans}}},\
  }\bibfield  {title} {\enquote {\bibinfo {title} {{Transient small-scale
  brightenings in the quiet solar corona: A model for campfires observed with
  Solar Orbiter}},}\ }\href {\doibase 10.1051/0004-6361/202140638} {\bibfield
  {journal} {\bibinfo  {journal} {Astronomy \& Astrophysics}\ }\textbf
  {\bibinfo {volume} {656}},\ \bibinfo {eid} {L7} (\bibinfo {year} {2021})},\
  \Eprint {http://arxiv.org/abs/2104.10940} {arXiv:2104.10940 [astro-ph.SR]}
  \BibitemShut {NoStop}%
\bibitem [{\citenamefont {Dong}\ \emph {et~al.}(2022)\citenamefont {Dong},
  \citenamefont {Wang}, \citenamefont {Huang}, \citenamefont {Comisso},
  \citenamefont {Sandstrom},\ and\ \citenamefont {Bhattacharjee}}]{Dong2022}%
  \BibitemOpen
  \bibfield  {author} {\bibinfo {author} {\bibfnamefont {C.}~\bibnamefont
  {Dong}}, \bibinfo {author} {\bibfnamefont {L.}~\bibnamefont {Wang}}, \bibinfo
  {author} {\bibfnamefont {Y.-M.}\ \bibnamefont {Huang}}, \bibinfo {author}
  {\bibfnamefont {L.}~\bibnamefont {Comisso}}, \bibinfo {author} {\bibfnamefont
  {T.~A.}\ \bibnamefont {Sandstrom}}, \ and\ \bibinfo {author} {\bibfnamefont
  {A.}~\bibnamefont {Bhattacharjee}},\ }\bibfield  {title} {\enquote {\bibinfo
  {title} {Reconnection-driven energy cascade in magnetohydrodynamic
  turbulence},}\ }\href {\doibase 10.1126/sciadv.abn7627} {\bibfield  {journal}
  {\bibinfo  {journal} {Science Advances}\ }\textbf {\bibinfo {volume} {8}},\
  \bibinfo {pages} {eabn7627} (\bibinfo {year} {2022})}\BibitemShut {NoStop}%
\bibitem [{\citenamefont {Qiu}\ \emph {et~al.}(2007)\citenamefont {Qiu},
  \citenamefont {Hu}, \citenamefont {Howard},\ and\ \citenamefont
  {Yurchyshyn}}]{Qiu2007}%
  \BibitemOpen
  \bibfield  {author} {\bibinfo {author} {\bibfnamefont {J.}~\bibnamefont
  {Qiu}}, \bibinfo {author} {\bibfnamefont {Q.}~\bibnamefont {Hu}}, \bibinfo
  {author} {\bibfnamefont {T.~A.}\ \bibnamefont {Howard}}, \ and\ \bibinfo
  {author} {\bibfnamefont {V.~B.}\ \bibnamefont {Yurchyshyn}},\ }\bibfield
  {title} {\enquote {\bibinfo {title} {On the magnetic flux budget in
  low-corona magnetic reconnection and interplanetary coronal mass
  ejections},}\ }\href@noop {} {\bibfield  {journal} {\bibinfo  {journal} {The
  Astrophysical Journal}\ }\textbf {\bibinfo {volume} {659}},\ \bibinfo {pages}
  {758–772} (\bibinfo {year} {2007})}\BibitemShut {NoStop}%
\bibitem [{\citenamefont {Chen}\ \emph {et~al.}(2008)\citenamefont {Chen},
  \citenamefont {Bhattacharjee}, \citenamefont {Puhl-Quinn}, \citenamefont
  {Yang}, \citenamefont {Bessho}, \citenamefont {Imada}, \citenamefont
  {Mühlbachler}, \citenamefont {Daly}, \citenamefont {Lefebvre}, \citenamefont
  {Khotyaintsev}, \citenamefont {Vaivads}, \citenamefont {Fazakerley},\ and\
  \citenamefont {Georgescu}}]{Chen2008}%
  \BibitemOpen
  \bibfield  {author} {\bibinfo {author} {\bibfnamefont {L.~J.}\ \bibnamefont
  {Chen}}, \bibinfo {author} {\bibfnamefont {A.}~\bibnamefont {Bhattacharjee}},
  \bibinfo {author} {\bibfnamefont {P.~A.}\ \bibnamefont {Puhl-Quinn}},
  \bibinfo {author} {\bibfnamefont {H.}~\bibnamefont {Yang}}, \bibinfo {author}
  {\bibfnamefont {N.}~\bibnamefont {Bessho}}, \bibinfo {author} {\bibfnamefont
  {S.}~\bibnamefont {Imada}}, \bibinfo {author} {\bibfnamefont
  {S.}~\bibnamefont {Mühlbachler}}, \bibinfo {author} {\bibfnamefont {P.~W.}\
  \bibnamefont {Daly}}, \bibinfo {author} {\bibfnamefont {B.}~\bibnamefont
  {Lefebvre}}, \bibinfo {author} {\bibfnamefont {Y.}~\bibnamefont
  {Khotyaintsev}}, \bibinfo {author} {\bibfnamefont {A.}~\bibnamefont
  {Vaivads}}, \bibinfo {author} {\bibfnamefont {A.}~\bibnamefont {Fazakerley}},
  \ and\ \bibinfo {author} {\bibfnamefont {E.}~\bibnamefont {Georgescu}},\
  }\bibfield  {title} {\enquote {\bibinfo {title} {Observation of energetic
  electrons within magnetic islands},}\ }\href@noop {} {\bibfield  {journal}
  {\bibinfo  {journal} {Nature Physics}\ }\textbf {\bibinfo {volume} {4}},\
  \bibinfo {pages} {19--23} (\bibinfo {year} {2008})}\BibitemShut {NoStop}%
\bibitem [{\citenamefont {{Le}}\ \emph {et~al.}(2017)\citenamefont {{Le}},
  \citenamefont {{Daughton}}, \citenamefont {{Chen}},\ and\ \citenamefont
  {{Egedal}}}]{Le2017}%
  \BibitemOpen
  \bibfield  {author} {\bibinfo {author} {\bibfnamefont {A.}~\bibnamefont
  {{Le}}}, \bibinfo {author} {\bibfnamefont {W.}~\bibnamefont {{Daughton}}},
  \bibinfo {author} {\bibfnamefont {L.~J.}\ \bibnamefont {{Chen}}}, \ and\
  \bibinfo {author} {\bibfnamefont {J.}~\bibnamefont {{Egedal}}},\ }\bibfield
  {title} {\enquote {\bibinfo {title} {{Enhanced electron mixing and heating in
  3-D asymmetric reconnection at the Earth's magnetopause}},}\ }\href {\doibase
  10.1002/2017GL072522} {\bibfield  {journal} {\bibinfo  {journal} {Geophysical
  Research Letters}\ }\textbf {\bibinfo {volume} {44}},\ \bibinfo {pages}
  {2096--2104} (\bibinfo {year} {2017})},\ \Eprint
  {http://arxiv.org/abs/1703.10246} {arXiv:1703.10246 [physics.plasm-ph]}
  \BibitemShut {NoStop}%
\bibitem [{\citenamefont {Phan}\ \emph {et~al.}(2018)\citenamefont {Phan},
  \citenamefont {Eastwood}, \citenamefont {Shay}, \citenamefont {Drake},
  \citenamefont {Sonnerup}, \citenamefont {Fujimoto}, \citenamefont {Cassak},
  \citenamefont {Øieroset}, \citenamefont {Burch}, \citenamefont {Torbert},
  \citenamefont {Rager}, \citenamefont {Dorelli}, \citenamefont {Gershman},
  \citenamefont {Pollock}, \citenamefont {Pyakurel}, \citenamefont {Haggerty},
  \citenamefont {Khotyaintsev}, \citenamefont {Lavraud}, \citenamefont {Saito},
  \citenamefont {Oka}, \citenamefont {Ergun}, \citenamefont {Retino},
  \citenamefont {Contel}, \citenamefont {Argall}, \citenamefont {Giles},
  \citenamefont {Moore}, \citenamefont {Wilder}, \citenamefont {Strangeway},
  \citenamefont {Russel}, \citenamefont {Lindqvist},\ and\ \citenamefont
  {Magnes}}]{Phan2018}%
  \BibitemOpen
  \bibfield  {author} {\bibinfo {author} {\bibfnamefont {T.~D.}\ \bibnamefont
  {Phan}}, \bibinfo {author} {\bibfnamefont {J.~P.}\ \bibnamefont {Eastwood}},
  \bibinfo {author} {\bibfnamefont {M.~A.}\ \bibnamefont {Shay}}, \bibinfo
  {author} {\bibfnamefont {J.~F.}\ \bibnamefont {Drake}}, \bibinfo {author}
  {\bibfnamefont {B.~U.~O.}\ \bibnamefont {Sonnerup}}, \bibinfo {author}
  {\bibfnamefont {M.}~\bibnamefont {Fujimoto}}, \bibinfo {author}
  {\bibfnamefont {P.~A.}\ \bibnamefont {Cassak}}, \bibinfo {author}
  {\bibfnamefont {M.}~\bibnamefont {Øieroset}}, \bibinfo {author}
  {\bibfnamefont {J.~L.}\ \bibnamefont {Burch}}, \bibinfo {author}
  {\bibfnamefont {R.~B.}\ \bibnamefont {Torbert}}, \bibinfo {author}
  {\bibfnamefont {A.~C.}\ \bibnamefont {Rager}}, \bibinfo {author}
  {\bibfnamefont {J.~C.}\ \bibnamefont {Dorelli}}, \bibinfo {author}
  {\bibfnamefont {D.~J.}\ \bibnamefont {Gershman}}, \bibinfo {author}
  {\bibfnamefont {C.}~\bibnamefont {Pollock}}, \bibinfo {author} {\bibfnamefont
  {P.~S.}\ \bibnamefont {Pyakurel}}, \bibinfo {author} {\bibfnamefont {C.~C.}\
  \bibnamefont {Haggerty}}, \bibinfo {author} {\bibfnamefont {Y.}~\bibnamefont
  {Khotyaintsev}}, \bibinfo {author} {\bibfnamefont {B.}~\bibnamefont
  {Lavraud}}, \bibinfo {author} {\bibfnamefont {Y.}~\bibnamefont {Saito}},
  \bibinfo {author} {\bibfnamefont {M.}~\bibnamefont {Oka}}, \bibinfo {author}
  {\bibfnamefont {R.~E.}\ \bibnamefont {Ergun}}, \bibinfo {author}
  {\bibfnamefont {A.}~\bibnamefont {Retino}}, \bibinfo {author} {\bibfnamefont
  {O.~L.}\ \bibnamefont {Contel}}, \bibinfo {author} {\bibfnamefont {M.~R.}\
  \bibnamefont {Argall}}, \bibinfo {author} {\bibfnamefont {B.~L.}\
  \bibnamefont {Giles}}, \bibinfo {author} {\bibfnamefont {T.~E.}\ \bibnamefont
  {Moore}}, \bibinfo {author} {\bibfnamefont {F.~D.}\ \bibnamefont {Wilder}},
  \bibinfo {author} {\bibfnamefont {R.~J.}\ \bibnamefont {Strangeway}},
  \bibinfo {author} {\bibfnamefont {C.~T.}\ \bibnamefont {Russel}}, \bibinfo
  {author} {\bibfnamefont {P.~A.}\ \bibnamefont {Lindqvist}}, \ and\ \bibinfo
  {author} {\bibfnamefont {W.}~\bibnamefont {Magnes}},\ }\bibfield  {title}
  {\enquote {\bibinfo {title} {Electron magnetic reconnection without ion
  coupling in earth’s turbulent magnetosheath},}\ }\href@noop {} {\bibfield
  {journal} {\bibinfo  {journal} {Nature}\ }\textbf {\bibinfo {volume} {557}},\
  \bibinfo {pages} {202--206} (\bibinfo {year} {2018})}\BibitemShut {NoStop}%
\bibitem [{\citenamefont {{Yang}}, \citenamefont {{Wang}},\ and\ \citenamefont
  {{Dong}}(2022)}]{Yang2022}%
  \BibitemOpen
  \bibfield  {author} {\bibinfo {author} {\bibfnamefont {S.-D.}\ \bibnamefont
  {{Yang}}}, \bibinfo {author} {\bibfnamefont {L.}~\bibnamefont {{Wang}}}, \
  and\ \bibinfo {author} {\bibfnamefont {C.}~\bibnamefont {{Dong}}},\
  }\bibfield  {title} {\enquote {\bibinfo {title} {{Discovery of Double Hall
  Pattern Associated with Collisionless Magnetic Reconnection in Dusty
  Plasmas}},}\ }\href@noop {} {\bibfield  {journal} {\bibinfo  {journal} {arXiv
  e-prints}\ ,\ \bibinfo {eid} {arXiv:2206.08553}} (\bibinfo {year} {2022})},\
  \Eprint {http://arxiv.org/abs/2206.08553} {arXiv:2206.08553
  [physics.plasm-ph]} \BibitemShut {NoStop}%
\bibitem [{\citenamefont {Yamada}\ \emph {et~al.}(1994)\citenamefont {Yamada},
  \citenamefont {Levinton}, \citenamefont {Pomphrey}, \citenamefont {Budny},
  \citenamefont {Manickam},\ and\ \citenamefont {Nagayama}}]{Yamada1994}%
  \BibitemOpen
  \bibfield  {author} {\bibinfo {author} {\bibfnamefont {M.}~\bibnamefont
  {Yamada}}, \bibinfo {author} {\bibfnamefont {F.~M.}\ \bibnamefont
  {Levinton}}, \bibinfo {author} {\bibfnamefont {N.}~\bibnamefont {Pomphrey}},
  \bibinfo {author} {\bibfnamefont {R.}~\bibnamefont {Budny}}, \bibinfo
  {author} {\bibfnamefont {J.}~\bibnamefont {Manickam}}, \ and\ \bibinfo
  {author} {\bibfnamefont {Y.}~\bibnamefont {Nagayama}},\ }\bibfield  {title}
  {\enquote {\bibinfo {title} {Investigation of magnetic reconnection during a
  sawtooth crash in a high‐temperature tokamak plasma},}\ }\href@noop {}
  {\bibfield  {journal} {\bibinfo  {journal} {Physics of Plasmas}\ }\textbf
  {\bibinfo {volume} {1}},\ \bibinfo {pages} {3269--3276} (\bibinfo {year}
  {1994})}\BibitemShut {NoStop}%
\bibitem [{\citenamefont {Duan}\ \emph {et~al.}(2018)\citenamefont {Duan},
  \citenamefont {Kan}, \citenamefont {Xiao}, \citenamefont {Xu}, \citenamefont
  {Yang}, \citenamefont {Wang}, \citenamefont {Huang},\ and\ \citenamefont
  {Xie}}]{Duan2018}%
  \BibitemOpen
  \bibfield  {author} {\bibinfo {author} {\bibfnamefont {S.~C.}\ \bibnamefont
  {Duan}}, \bibinfo {author} {\bibfnamefont {M.~X.}\ \bibnamefont {Kan}},
  \bibinfo {author} {\bibfnamefont {B.}~\bibnamefont {Xiao}}, \bibinfo {author}
  {\bibfnamefont {Q.}~\bibnamefont {Xu}}, \bibinfo {author} {\bibfnamefont
  {L.}~\bibnamefont {Yang}}, \bibinfo {author} {\bibfnamefont {G.~H.}\
  \bibnamefont {Wang}}, \bibinfo {author} {\bibfnamefont {X.~B.}\ \bibnamefont
  {Huang}}, \ and\ \bibinfo {author} {\bibfnamefont {W.~P.}\ \bibnamefont
  {Xie}},\ }\bibfield  {title} {\enquote {\bibinfo {title} {Numerical modelling
  of inverse wire array z-pinch magnetic reconnection},}\ }\href@noop {}
  {\bibfield  {journal} {\bibinfo  {journal} {AIP Advances}\ }\textbf {\bibinfo
  {volume} {8}},\ \bibinfo {pages} {055018} (\bibinfo {year}
  {2018})}\BibitemShut {NoStop}%
\bibitem [{\citenamefont {{Raymond}}\ \emph {et~al.}(2018)\citenamefont
  {{Raymond}}, \citenamefont {{Dong}}, \citenamefont {{McKelvey}},
  \citenamefont {{Zulick}}, \citenamefont {{Alexander}}, \citenamefont
  {{Bhattacharjee}}, \citenamefont {{Campbell}}, \citenamefont {{Chen}},
  \citenamefont {{Chvykov}}, \citenamefont {{Del Rio}}, \citenamefont
  {{Fitzsimmons}}, \citenamefont {{Fox}}, \citenamefont {{Hou}}, \citenamefont
  {{Maksimchuk}}, \citenamefont {{Mileham}}, \citenamefont {{Nees}},
  \citenamefont {{Nilson}}, \citenamefont {{Stoeckl}}, \citenamefont
  {{Thomas}}, \citenamefont {{Wei}}, \citenamefont {{Yanovsky}}, \citenamefont
  {{Krushelnick}},\ and\ \citenamefont {{Willingale}}}]{Raymond2018}%
  \BibitemOpen
  \bibfield  {author} {\bibinfo {author} {\bibfnamefont {A.~E.}\ \bibnamefont
  {{Raymond}}}, \bibinfo {author} {\bibfnamefont {C.~F.}\ \bibnamefont
  {{Dong}}}, \bibinfo {author} {\bibfnamefont {A.}~\bibnamefont {{McKelvey}}},
  \bibinfo {author} {\bibfnamefont {C.}~\bibnamefont {{Zulick}}}, \bibinfo
  {author} {\bibfnamefont {N.}~\bibnamefont {{Alexander}}}, \bibinfo {author}
  {\bibfnamefont {A.}~\bibnamefont {{Bhattacharjee}}}, \bibinfo {author}
  {\bibfnamefont {P.~T.}\ \bibnamefont {{Campbell}}}, \bibinfo {author}
  {\bibfnamefont {H.}~\bibnamefont {{Chen}}}, \bibinfo {author} {\bibfnamefont
  {V.}~\bibnamefont {{Chvykov}}}, \bibinfo {author} {\bibfnamefont
  {E.}~\bibnamefont {{Del Rio}}}, \bibinfo {author} {\bibfnamefont
  {P.}~\bibnamefont {{Fitzsimmons}}}, \bibinfo {author} {\bibfnamefont
  {W.}~\bibnamefont {{Fox}}}, \bibinfo {author} {\bibfnamefont
  {B.}~\bibnamefont {{Hou}}}, \bibinfo {author} {\bibfnamefont
  {A.}~\bibnamefont {{Maksimchuk}}}, \bibinfo {author} {\bibfnamefont
  {C.}~\bibnamefont {{Mileham}}}, \bibinfo {author} {\bibfnamefont
  {J.}~\bibnamefont {{Nees}}}, \bibinfo {author} {\bibfnamefont {P.~M.}\
  \bibnamefont {{Nilson}}}, \bibinfo {author} {\bibfnamefont {C.}~\bibnamefont
  {{Stoeckl}}}, \bibinfo {author} {\bibfnamefont {A.~G.~R.}\ \bibnamefont
  {{Thomas}}}, \bibinfo {author} {\bibfnamefont {M.~S.}\ \bibnamefont {{Wei}}},
  \bibinfo {author} {\bibfnamefont {V.}~\bibnamefont {{Yanovsky}}}, \bibinfo
  {author} {\bibfnamefont {K.}~\bibnamefont {{Krushelnick}}}, \ and\ \bibinfo
  {author} {\bibfnamefont {L.}~\bibnamefont {{Willingale}}},\ }\bibfield
  {title} {\enquote {\bibinfo {title} {{Relativistic-electron-driven magnetic
  reconnection in the laboratory}},}\ }\href {\doibase
  10.1103/PhysRevE.98.043207} {\bibfield  {journal} {\bibinfo  {journal}
  {Physical Review E}\ }\textbf {\bibinfo {volume} {98}},\ \bibinfo {eid}
  {043207} (\bibinfo {year} {2018})}\BibitemShut {NoStop}%
\bibitem [{\citenamefont {Ji}\ \emph {et~al.}(2022)\citenamefont {Ji},
  \citenamefont {Daughton}, \citenamefont {Jara-Almonte}, \citenamefont {Le},
  \citenamefont {Stanier},\ and\ \citenamefont {Yoo}}]{Ji2022}%
  \BibitemOpen
  \bibfield  {author} {\bibinfo {author} {\bibfnamefont {H.}~\bibnamefont
  {Ji}}, \bibinfo {author} {\bibfnamefont {W.}~\bibnamefont {Daughton}},
  \bibinfo {author} {\bibfnamefont {J.}~\bibnamefont {Jara-Almonte}}, \bibinfo
  {author} {\bibfnamefont {A.}~\bibnamefont {Le}}, \bibinfo {author}
  {\bibfnamefont {A.}~\bibnamefont {Stanier}}, \ and\ \bibinfo {author}
  {\bibfnamefont {J.}~\bibnamefont {Yoo}},\ }\bibfield  {title} {\enquote
  {\bibinfo {title} {Magnetic reconnection in the era of exascale computing and
  multiscale experiments},}\ }\href@noop {} {\bibfield  {journal} {\bibinfo
  {journal} {Nature Reviews Physics}\ }\textbf {\bibinfo {volume} {4}},\
  \bibinfo {pages} {263--282} (\bibinfo {year} {2022})}\BibitemShut {NoStop}%
\bibitem [{\citenamefont {{Zweibel}}\ and\ \citenamefont
  {{Yamada}}(2009)}]{Zweibel2009}%
  \BibitemOpen
  \bibfield  {author} {\bibinfo {author} {\bibfnamefont {E.~G.}\ \bibnamefont
  {{Zweibel}}}\ and\ \bibinfo {author} {\bibfnamefont {M.}~\bibnamefont
  {{Yamada}}},\ }\bibfield  {title} {\enquote {\bibinfo {title} {Magnetic
  reconnection in astrophysical and laboratory plasmas},}\ }\href {\doibase
  10.1146/annurev-astro-082708-101726} {\bibfield  {journal} {\bibinfo
  {journal} {Annual Review of Astronomy and Astrophysics}\ }\textbf {\bibinfo
  {volume} {47}},\ \bibinfo {pages} {291--332} (\bibinfo {year}
  {2009})}\BibitemShut {NoStop}%
\bibitem [{\citenamefont {Yamada}, \citenamefont {Kulsrud},\ and\ \citenamefont
  {Ji}(2010)}]{Yamada2010}%
  \BibitemOpen
  \bibfield  {author} {\bibinfo {author} {\bibfnamefont {M.}~\bibnamefont
  {Yamada}}, \bibinfo {author} {\bibfnamefont {R.}~\bibnamefont {Kulsrud}}, \
  and\ \bibinfo {author} {\bibfnamefont {H.}~\bibnamefont {Ji}},\ }\bibfield
  {title} {\enquote {\bibinfo {title} {Magnetic reconnection},}\ }\href
  {\doibase 10.1103/RevModPhys.82.603} {\bibfield  {journal} {\bibinfo
  {journal} {Rev. Mod. Phys.}\ }\textbf {\bibinfo {volume} {82}},\ \bibinfo
  {pages} {603--664} (\bibinfo {year} {2010})}\BibitemShut {NoStop}%
\bibitem [{\citenamefont {{Daldorff}}\ \emph {et~al.}(2014)\citenamefont
  {{Daldorff}}, \citenamefont {{T{\'o}th}}, \citenamefont {{Gombosi}},
  \citenamefont {{Lapenta}}, \citenamefont {{Amaya}}, \citenamefont
  {{Markidis}},\ and\ \citenamefont {{Brackbill}}}]{Daldorff2014}%
  \BibitemOpen
  \bibfield  {author} {\bibinfo {author} {\bibfnamefont {L.~K.~S.}\
  \bibnamefont {{Daldorff}}}, \bibinfo {author} {\bibfnamefont
  {G.}~\bibnamefont {{T{\'o}th}}}, \bibinfo {author} {\bibfnamefont {T.~I.}\
  \bibnamefont {{Gombosi}}}, \bibinfo {author} {\bibfnamefont {G.}~\bibnamefont
  {{Lapenta}}}, \bibinfo {author} {\bibfnamefont {J.}~\bibnamefont {{Amaya}}},
  \bibinfo {author} {\bibfnamefont {S.}~\bibnamefont {{Markidis}}}, \ and\
  \bibinfo {author} {\bibfnamefont {J.~U.}\ \bibnamefont {{Brackbill}}},\
  }\bibfield  {title} {\enquote {\bibinfo {title} {{Two-way coupling of a
  global Hall magnetohydrodynamics model with a local implicit particle-in-cell
  model}},}\ }\href {\doibase 10.1016/j.jcp.2014.03.009} {\bibfield  {journal}
  {\bibinfo  {journal} {Journal of Computational Physics}\ }\textbf {\bibinfo
  {volume} {268}},\ \bibinfo {pages} {236--254} (\bibinfo {year}
  {2014})}\BibitemShut {NoStop}%
\bibitem [{\citenamefont {{T{\'o}th}}\ \emph {et~al.}(2016)\citenamefont
  {{T{\'o}th}}, \citenamefont {{Jia}}, \citenamefont {{Markidis}},
  \citenamefont {{Peng}}, \citenamefont {{Chen}}, \citenamefont {{Daldorff}},
  \citenamefont {{Tenishev}}, \citenamefont {{Borovikov}}, \citenamefont
  {{Haiducek}}, \citenamefont {{Gombosi}}, \citenamefont {{Glocer}},\ and\
  \citenamefont {{Dorelli}}}]{Toth2016}%
  \BibitemOpen
  \bibfield  {author} {\bibinfo {author} {\bibfnamefont {G.}~\bibnamefont
  {{T{\'o}th}}}, \bibinfo {author} {\bibfnamefont {X.}~\bibnamefont {{Jia}}},
  \bibinfo {author} {\bibfnamefont {S.}~\bibnamefont {{Markidis}}}, \bibinfo
  {author} {\bibfnamefont {I.~B.}\ \bibnamefont {{Peng}}}, \bibinfo {author}
  {\bibfnamefont {Y.}~\bibnamefont {{Chen}}}, \bibinfo {author} {\bibfnamefont
  {L.~K.~S.}\ \bibnamefont {{Daldorff}}}, \bibinfo {author} {\bibfnamefont
  {V.~M.}\ \bibnamefont {{Tenishev}}}, \bibinfo {author} {\bibfnamefont
  {D.}~\bibnamefont {{Borovikov}}}, \bibinfo {author} {\bibfnamefont {J.~D.}\
  \bibnamefont {{Haiducek}}}, \bibinfo {author} {\bibfnamefont {T.~I.}\
  \bibnamefont {{Gombosi}}}, \bibinfo {author} {\bibfnamefont {A.}~\bibnamefont
  {{Glocer}}}, \ and\ \bibinfo {author} {\bibfnamefont {J.~C.}\ \bibnamefont
  {{Dorelli}}},\ }\bibfield  {title} {\enquote {\bibinfo {title} {{Extended
  magnetohydrodynamics with embedded particle-in-cell simulation of Ganymede's
  magnetosphere}},}\ }\href {\doibase 10.1002/2015JA021997} {\bibfield
  {journal} {\bibinfo  {journal} {Journal of Geophysical Research (Space
  Physics)}\ }\textbf {\bibinfo {volume} {121}},\ \bibinfo {pages} {1273--1293}
  (\bibinfo {year} {2016})}\BibitemShut {NoStop}%
\bibitem [{\citenamefont {Chen}\ \emph {et~al.}(2017)\citenamefont {Chen},
  \citenamefont {Tóth}, \citenamefont {Cassak}, \citenamefont {Jia},
  \citenamefont {Gombosi}, \citenamefont {Slavin}, \citenamefont {Markidis},
  \citenamefont {Peng}, \citenamefont {Jordanova},\ and\ \citenamefont
  {Henderson}}]{Chen2017}%
  \BibitemOpen
  \bibfield  {author} {\bibinfo {author} {\bibfnamefont {Y.}~\bibnamefont
  {Chen}}, \bibinfo {author} {\bibfnamefont {G.}~\bibnamefont {Tóth}},
  \bibinfo {author} {\bibfnamefont {P.}~\bibnamefont {Cassak}}, \bibinfo
  {author} {\bibfnamefont {X.}~\bibnamefont {Jia}}, \bibinfo {author}
  {\bibfnamefont {T.~I.}\ \bibnamefont {Gombosi}}, \bibinfo {author}
  {\bibfnamefont {J.~A.}\ \bibnamefont {Slavin}}, \bibinfo {author}
  {\bibfnamefont {S.}~\bibnamefont {Markidis}}, \bibinfo {author}
  {\bibfnamefont {I.~B.}\ \bibnamefont {Peng}}, \bibinfo {author}
  {\bibfnamefont {V.~K.}\ \bibnamefont {Jordanova}}, \ and\ \bibinfo {author}
  {\bibfnamefont {M.~G.}\ \bibnamefont {Henderson}},\ }\bibfield  {title}
  {\enquote {\bibinfo {title} {Global three-dimensional simulation of earth's
  dayside reconnection using a two-way coupled magnetohydrodynamics with
  embedded particle-in-cell model: Initial results},}\ }\href@noop {}
  {\bibfield  {journal} {\bibinfo  {journal} {Journal of Geophysical Research:
  Space Physics}\ }\textbf {\bibinfo {volume} {122}},\ \bibinfo {pages}
  {10,318--10,335} (\bibinfo {year} {2017})},\ \bibinfo {note}
  {{DOI:10.1002/2017JA024186}}\BibitemShut {NoStop}%
\bibitem [{\citenamefont {{Chen}}\ \emph {et~al.}(2019)\citenamefont {{Chen}},
  \citenamefont {{T{\'o}th}}, \citenamefont {{Jia}}, \citenamefont {{Slavin}},
  \citenamefont {{Sun}}, \citenamefont {{Markidis}}, \citenamefont
  {{Gombosi}},\ and\ \citenamefont {{Raines}}}]{Chen2019JGR}%
  \BibitemOpen
  \bibfield  {author} {\bibinfo {author} {\bibfnamefont {Y.}~\bibnamefont
  {{Chen}}}, \bibinfo {author} {\bibfnamefont {G.}~\bibnamefont {{T{\'o}th}}},
  \bibinfo {author} {\bibfnamefont {X.}~\bibnamefont {{Jia}}}, \bibinfo
  {author} {\bibfnamefont {J.~A.}\ \bibnamefont {{Slavin}}}, \bibinfo {author}
  {\bibfnamefont {W.}~\bibnamefont {{Sun}}}, \bibinfo {author} {\bibfnamefont
  {S.}~\bibnamefont {{Markidis}}}, \bibinfo {author} {\bibfnamefont {T.~I.}\
  \bibnamefont {{Gombosi}}}, \ and\ \bibinfo {author} {\bibfnamefont {J.~M.}\
  \bibnamefont {{Raines}}},\ }\bibfield  {title} {\enquote {\bibinfo {title}
  {{Studying Dawn-Dusk Asymmetries of Mercury's Magnetotail Using MHD-EPIC
  Simulations}},}\ }\href {\doibase 10.1029/2019JA026840} {\bibfield  {journal}
  {\bibinfo  {journal} {Journal of Geophysical Research (Space Physics)}\
  }\textbf {\bibinfo {volume} {124}},\ \bibinfo {pages} {8954--8973} (\bibinfo
  {year} {2019})},\ \Eprint {http://arxiv.org/abs/1904.06753} {arXiv:1904.06753
  [physics.space-ph]} \BibitemShut {NoStop}%
\bibitem [{\citenamefont {Wang}\ \emph {et~al.}(2015)\citenamefont {Wang},
  \citenamefont {Hakim}, \citenamefont {Bhattacharjee},\ and\ \citenamefont
  {Germaschewski}}]{Wang2015}%
  \BibitemOpen
  \bibfield  {author} {\bibinfo {author} {\bibfnamefont {L.}~\bibnamefont
  {Wang}}, \bibinfo {author} {\bibfnamefont {A.~H.}\ \bibnamefont {Hakim}},
  \bibinfo {author} {\bibfnamefont {A.}~\bibnamefont {Bhattacharjee}}, \ and\
  \bibinfo {author} {\bibfnamefont {K.}~\bibnamefont {Germaschewski}},\
  }\bibfield  {title} {\enquote {\bibinfo {title} {Comparison of multi-fluid
  moment models with particle-in-cell simulations of collisionless magnetic
  reconnection},}\ }\href@noop {} {\bibfield  {journal} {\bibinfo  {journal}
  {Physics of Plasmas}\ }\textbf {\bibinfo {volume} {22}},\ \bibinfo {pages}
  {012108} (\bibinfo {year} {2015})},\ \bibinfo {note}
  {{DOI:10.1063/1.4906063}}\BibitemShut {NoStop}%
\bibitem [{\citenamefont {{Wang}}\ \emph {et~al.}(2018)\citenamefont {{Wang}},
  \citenamefont {{Germaschewski}}, \citenamefont {{Hakim}}, \citenamefont
  {{Dong}}, \citenamefont {{Raeder}},\ and\ \citenamefont
  {{Bhattacharjee}}}]{Wang2018}%
  \BibitemOpen
  \bibfield  {author} {\bibinfo {author} {\bibfnamefont {L.}~\bibnamefont
  {{Wang}}}, \bibinfo {author} {\bibfnamefont {K.}~\bibnamefont
  {{Germaschewski}}}, \bibinfo {author} {\bibfnamefont {A.}~\bibnamefont
  {{Hakim}}}, \bibinfo {author} {\bibfnamefont {C.}~\bibnamefont {{Dong}}},
  \bibinfo {author} {\bibfnamefont {J.}~\bibnamefont {{Raeder}}}, \ and\
  \bibinfo {author} {\bibfnamefont {A.}~\bibnamefont {{Bhattacharjee}}},\
  }\bibfield  {title} {\enquote {\bibinfo {title} {{Electron Physics in 3-D
  Two-Fluid 10-Moment Modeling of Ganymede's Magnetosphere}},}\ }\href@noop {}
  {\bibfield  {journal} {\bibinfo  {journal} {Journal of Geophysical Research
  (Space Physics)}\ }\textbf {\bibinfo {volume} {123}},\ \bibinfo {pages}
  {2815--2830} (\bibinfo {year} {2018})}\BibitemShut {NoStop}%
\bibitem [{\citenamefont {{Dong}}\ \emph {et~al.}(2019)\citenamefont {{Dong}},
  \citenamefont {{Wang}}, \citenamefont {{Hakim}}, \citenamefont
  {{Bhattacharjee}}, \citenamefont {{Slavin}}, \citenamefont {{DiBraccio}},\
  and\ \citenamefont {{Germaschewski}}}]{Dong2019}%
  \BibitemOpen
  \bibfield  {author} {\bibinfo {author} {\bibfnamefont {C.}~\bibnamefont
  {{Dong}}}, \bibinfo {author} {\bibfnamefont {L.}~\bibnamefont {{Wang}}},
  \bibinfo {author} {\bibfnamefont {A.}~\bibnamefont {{Hakim}}}, \bibinfo
  {author} {\bibfnamefont {A.}~\bibnamefont {{Bhattacharjee}}}, \bibinfo
  {author} {\bibfnamefont {J.~A.}\ \bibnamefont {{Slavin}}}, \bibinfo {author}
  {\bibfnamefont {G.~A.}\ \bibnamefont {{DiBraccio}}}, \ and\ \bibinfo {author}
  {\bibfnamefont {K.}~\bibnamefont {{Germaschewski}}},\ }\bibfield  {title}
  {\enquote {\bibinfo {title} {{Global Ten-Moment Multifluid Simulations of the
  Solar Wind Interaction with Mercury: From the Planetary Conducting Core to
  the Dynamic Magnetosphere}},}\ }\href@noop {} {\bibfield  {journal} {\bibinfo
   {journal} {Geophysical Research Letters}\ }\textbf {\bibinfo {volume}
  {46}},\ \bibinfo {pages} {11,584--11,596} (\bibinfo {year}
  {2019})}\BibitemShut {NoStop}%
\bibitem [{\citenamefont {{Wang}}\ \emph {et~al.}(2020)\citenamefont {{Wang}},
  \citenamefont {{Hakim}}, \citenamefont {{Ng}}, \citenamefont {{Dong}},\ and\
  \citenamefont {{Germaschewski}}}]{Wang2020}%
  \BibitemOpen
  \bibfield  {author} {\bibinfo {author} {\bibfnamefont {L.}~\bibnamefont
  {{Wang}}}, \bibinfo {author} {\bibfnamefont {A.~H.}\ \bibnamefont {{Hakim}}},
  \bibinfo {author} {\bibfnamefont {J.}~\bibnamefont {{Ng}}}, \bibinfo {author}
  {\bibfnamefont {C.}~\bibnamefont {{Dong}}}, \ and\ \bibinfo {author}
  {\bibfnamefont {K.}~\bibnamefont {{Germaschewski}}},\ }\bibfield  {title}
  {\enquote {\bibinfo {title} {{Exact and locally implicit source term solvers
  for multifluid-Maxwell systems}},}\ }\href@noop {} {\bibfield  {journal}
  {\bibinfo  {journal} {Journal of Computational Physics}\ }\textbf {\bibinfo
  {volume} {415}},\ \bibinfo {eid} {109510} (\bibinfo {year}
  {2020})}\BibitemShut {NoStop}%
\bibitem [{\citenamefont {{Jarmak}}\ \emph {et~al.}(2020)\citenamefont
  {{Jarmak}}, \citenamefont {{Leonard}}, \citenamefont {{Akins}}, \citenamefont
  {{Dahl}}, \citenamefont {{Cremons}}, \citenamefont {{Cofield}}, \citenamefont
  {{Curtis}}, \citenamefont {{Dong}}, \citenamefont {{Dunham}}, \citenamefont
  {{Journaux}}, \citenamefont {{Murakami}}, \citenamefont {{Ng}}, \citenamefont
  {{Piquette}}, \citenamefont {{Girija}}, \citenamefont {{Rink}}, \citenamefont
  {{Schurmeier}}, \citenamefont {{Stein}}, \citenamefont {{Tallarida}},
  \citenamefont {{Telus}}, \citenamefont {{Lowes}}, \citenamefont {{Budney}},\
  and\ \citenamefont {{Mitchell}}}]{Jarmak2020}%
  \BibitemOpen
  \bibfield  {author} {\bibinfo {author} {\bibfnamefont {S.}~\bibnamefont
  {{Jarmak}}}, \bibinfo {author} {\bibfnamefont {E.}~\bibnamefont {{Leonard}}},
  \bibinfo {author} {\bibfnamefont {A.}~\bibnamefont {{Akins}}}, \bibinfo
  {author} {\bibfnamefont {E.}~\bibnamefont {{Dahl}}}, \bibinfo {author}
  {\bibfnamefont {D.~R.}\ \bibnamefont {{Cremons}}}, \bibinfo {author}
  {\bibfnamefont {S.}~\bibnamefont {{Cofield}}}, \bibinfo {author}
  {\bibfnamefont {A.}~\bibnamefont {{Curtis}}}, \bibinfo {author}
  {\bibfnamefont {C.}~\bibnamefont {{Dong}}}, \bibinfo {author} {\bibfnamefont
  {E.~T.}\ \bibnamefont {{Dunham}}}, \bibinfo {author} {\bibfnamefont
  {B.}~\bibnamefont {{Journaux}}}, \bibinfo {author} {\bibfnamefont
  {D.}~\bibnamefont {{Murakami}}}, \bibinfo {author} {\bibfnamefont
  {W.}~\bibnamefont {{Ng}}}, \bibinfo {author} {\bibfnamefont {M.}~\bibnamefont
  {{Piquette}}}, \bibinfo {author} {\bibfnamefont {A.~P.}\ \bibnamefont
  {{Girija}}}, \bibinfo {author} {\bibfnamefont {K.}~\bibnamefont {{Rink}}},
  \bibinfo {author} {\bibfnamefont {L.}~\bibnamefont {{Schurmeier}}}, \bibinfo
  {author} {\bibfnamefont {N.}~\bibnamefont {{Stein}}}, \bibinfo {author}
  {\bibfnamefont {N.}~\bibnamefont {{Tallarida}}}, \bibinfo {author}
  {\bibfnamefont {M.}~\bibnamefont {{Telus}}}, \bibinfo {author} {\bibfnamefont
  {L.}~\bibnamefont {{Lowes}}}, \bibinfo {author} {\bibfnamefont
  {C.}~\bibnamefont {{Budney}}}, \ and\ \bibinfo {author} {\bibfnamefont
  {K.~L.}\ \bibnamefont {{Mitchell}}},\ }\bibfield  {title} {\enquote {\bibinfo
  {title} {{QUEST: A New Frontiers Uranus orbiter mission concept study}},}\
  }\href@noop {} {\bibfield  {journal} {\bibinfo  {journal} {Acta
  Astronautica}\ }\textbf {\bibinfo {volume} {170}},\ \bibinfo {pages} {6--26}
  (\bibinfo {year} {2020})}\BibitemShut {NoStop}%
\bibitem [{\citenamefont {{Rulke}}, \citenamefont {{Wang}},\ and\ \citenamefont
  {{Dong}}(2021)}]{Rulke2021}%
  \BibitemOpen
  \bibfield  {author} {\bibinfo {author} {\bibfnamefont {E.}~\bibnamefont
  {{Rulke}}}, \bibinfo {author} {\bibfnamefont {L.}~\bibnamefont {{Wang}}}, \
  and\ \bibinfo {author} {\bibfnamefont {C.}~\bibnamefont {{Dong}}},\
  }\bibfield  {title} {\enquote {\bibinfo {title} {{Three-dimensional,
  ten-moment multifluid simulation of the solar wind interaction with Asteroid
  16 Psyche}},}\ }in\ \href@noop {} {\emph {\bibinfo {booktitle} {AGU Fall
  Meeting Abstracts}}},\ Vol.\ \bibinfo {volume} {2021}\ (\bibinfo {year}
  {2021})\ pp.\ \bibinfo {pages} {SM53C--08}\BibitemShut {NoStop}%
\bibitem [{\citenamefont {{Qin}}\ \emph {et~al.}(2022)\citenamefont {{Qin}},
  \citenamefont {{Ma}}, \citenamefont {{Jiang}}, \citenamefont {{Dong}},
  \citenamefont {{Fu}}, \citenamefont {{Wang}}, \citenamefont {{Cheng}},\ and\
  \citenamefont {{Jin}}}]{Qin2022}%
  \BibitemOpen
  \bibfield  {author} {\bibinfo {author} {\bibfnamefont {Y.}~\bibnamefont
  {{Qin}}}, \bibinfo {author} {\bibfnamefont {J.}~\bibnamefont {{Ma}}},
  \bibinfo {author} {\bibfnamefont {M.}~\bibnamefont {{Jiang}}}, \bibinfo
  {author} {\bibfnamefont {C.}~\bibnamefont {{Dong}}}, \bibinfo {author}
  {\bibfnamefont {H.}~\bibnamefont {{Fu}}}, \bibinfo {author} {\bibfnamefont
  {L.}~\bibnamefont {{Wang}}}, \bibinfo {author} {\bibfnamefont
  {W.}~\bibnamefont {{Cheng}}}, \ and\ \bibinfo {author} {\bibfnamefont
  {Y.}~\bibnamefont {{Jin}}},\ }\bibfield  {title} {\enquote {\bibinfo {title}
  {{Data-Driven Modeling of Landau Damping by Physics-Informed Neural
  Networks}},}\ }\href@noop {} {\bibfield  {journal} {\bibinfo  {journal}
  {arXiv e-prints}\ ,\ \bibinfo {eid} {arXiv:2211.01021}} (\bibinfo {year}
  {2022})},\ \Eprint {http://arxiv.org/abs/2211.01021} {arXiv:2211.01021
  [physics.plasm-ph]} \BibitemShut {NoStop}%
\bibitem [{\citenamefont {Cheng}\ \emph {et~al.}(2023)\citenamefont {Cheng},
  \citenamefont {Fu}, \citenamefont {Wang}, \citenamefont {Dong}, \citenamefont
  {Jin}, \citenamefont {Jiang}, \citenamefont {Ma}, \citenamefont {Qin},\ and\
  \citenamefont {Liu}}]{Cheng2023}%
  \BibitemOpen
  \bibfield  {author} {\bibinfo {author} {\bibfnamefont {W.}~\bibnamefont
  {Cheng}}, \bibinfo {author} {\bibfnamefont {H.}~\bibnamefont {Fu}}, \bibinfo
  {author} {\bibfnamefont {L.}~\bibnamefont {Wang}}, \bibinfo {author}
  {\bibfnamefont {C.}~\bibnamefont {Dong}}, \bibinfo {author} {\bibfnamefont
  {Y.}~\bibnamefont {Jin}}, \bibinfo {author} {\bibfnamefont {M.}~\bibnamefont
  {Jiang}}, \bibinfo {author} {\bibfnamefont {J.}~\bibnamefont {Ma}}, \bibinfo
  {author} {\bibfnamefont {Y.}~\bibnamefont {Qin}}, \ and\ \bibinfo {author}
  {\bibfnamefont {K.}~\bibnamefont {Liu}},\ }\bibfield  {title} {\enquote
  {\bibinfo {title} {Data-driven, multi-moment fluid modeling of landau
  damping},}\ }\href {\doibase https://doi.org/10.1016/j.cpc.2022.108538}
  {\bibfield  {journal} {\bibinfo  {journal} {Computer Physics Communications}\
  }\textbf {\bibinfo {volume} {282}},\ \bibinfo {pages} {108538} (\bibinfo
  {year} {2023})}\BibitemShut {NoStop}%
\bibitem [{\citenamefont {Shou}\ \emph {et~al.}(2021)\citenamefont {Shou},
  \citenamefont {Tenishev}, \citenamefont {Chen}, \citenamefont {T\'oth},\ and\
  \citenamefont {Ganushkina}}]{Shou2021}%
  \BibitemOpen
  \bibfield  {author} {\bibinfo {author} {\bibfnamefont {Y.}~\bibnamefont
  {Shou}}, \bibinfo {author} {\bibfnamefont {V.}~\bibnamefont {Tenishev}},
  \bibinfo {author} {\bibfnamefont {Y.}~\bibnamefont {Chen}}, \bibinfo {author}
  {\bibfnamefont {G.}~\bibnamefont {T\'oth}}, \ and\ \bibinfo {author}
  {\bibfnamefont {N.}~\bibnamefont {Ganushkina}},\ }\bibfield  {title}
  {\enquote {\bibinfo {title} {{Magnetohydrodynamic with Adaptively Embedded
  Particle-in-Cell model: MHD-AEPIC}},}\ }\href {\doibase
  https://doi.org/10.1016/j.jcp.2021.110656} {\bibfield  {journal} {\bibinfo
  {journal} {Journal of Computational Physics}\ }\textbf {\bibinfo {volume}
  {446}},\ \bibinfo {pages} {110656} (\bibinfo {year} {2021})}\BibitemShut
  {NoStop}%
\bibitem [{\citenamefont {Chen}\ \emph {et~al.}(2021)\citenamefont {Chen},
  \citenamefont {Tóth}, \citenamefont {Zhou},\ and\ \citenamefont
  {Wang}}]{Chen2021}%
  \BibitemOpen
  \bibfield  {author} {\bibinfo {author} {\bibfnamefont {Y.}~\bibnamefont
  {Chen}}, \bibinfo {author} {\bibfnamefont {G.}~\bibnamefont {Tóth}},
  \bibinfo {author} {\bibfnamefont {H.}~\bibnamefont {Zhou}}, \ and\ \bibinfo
  {author} {\bibfnamefont {X.}~\bibnamefont {Wang}},\ }\bibfield  {title}
  {\enquote {\bibinfo {title} {Fleks: A flexible particle-in-cell code for
  multi-scale plasma simulations},}\ }\href@noop {} {\bibfield  {journal}
  {\bibinfo  {journal} {Earth and Space Science Open Archive}\ } (\bibinfo
  {year} {2021})}\BibitemShut {NoStop}%
\bibitem [{\citenamefont {Wang}, \citenamefont {Chen},\ and\ \citenamefont
  {Tóth}(2022)}]{Wang2022}%
  \BibitemOpen
  \bibfield  {author} {\bibinfo {author} {\bibfnamefont {X.}~\bibnamefont
  {Wang}}, \bibinfo {author} {\bibfnamefont {Y.}~\bibnamefont {Chen}}, \ and\
  \bibinfo {author} {\bibfnamefont {G.}~\bibnamefont {Tóth}},\ }\bibfield
  {title} {\enquote {\bibinfo {title} {Global magnetohydrodynamic magnetosphere
  simulation with an adaptively embedded particle-in-cell model},}\ }\href
  {\doibase https://doi.org/10.1029/2021JA030091} {\bibfield  {journal}
  {\bibinfo  {journal} {Journal of Geophysical Research: Space Physics}\
  }\textbf {\bibinfo {volume} {127}},\ \bibinfo {pages} {e2021JA030091}
  (\bibinfo {year} {2022})}\BibitemShut {NoStop}%
\bibitem [{\citenamefont {Ng}\ \emph {et~al.}(2015)\citenamefont {Ng},
  \citenamefont {Huang}, \citenamefont {Hakim}, \citenamefont {Bhattacharjee},
  \citenamefont {Stanier}, \citenamefont {Daughton}, \citenamefont {Wang},\
  and\ \citenamefont {Germaschewski}}]{Ng2015}%
  \BibitemOpen
  \bibfield  {author} {\bibinfo {author} {\bibfnamefont {J.}~\bibnamefont
  {Ng}}, \bibinfo {author} {\bibfnamefont {Y.~M.}\ \bibnamefont {Huang}},
  \bibinfo {author} {\bibfnamefont {A.}~\bibnamefont {Hakim}}, \bibinfo
  {author} {\bibfnamefont {A.}~\bibnamefont {Bhattacharjee}}, \bibinfo {author}
  {\bibfnamefont {A.}~\bibnamefont {Stanier}}, \bibinfo {author} {\bibfnamefont
  {W.}~\bibnamefont {Daughton}}, \bibinfo {author} {\bibfnamefont
  {L.}~\bibnamefont {Wang}}, \ and\ \bibinfo {author} {\bibfnamefont
  {K.}~\bibnamefont {Germaschewski}},\ }\bibfield  {title} {\enquote {\bibinfo
  {title} {The island coalescence problem: Scaling of reconnection in extended
  fluid models including higher-order moments},}\ }\href@noop {} {\bibfield
  {journal} {\bibinfo  {journal} {Physics of Plasmas}\ }\textbf {\bibinfo
  {volume} {22}},\ \bibinfo {pages} {112104} (\bibinfo {year} {2015})},\
  \bibinfo {note} {{DOI:10.1063/1.4935302}}\BibitemShut {NoStop}%
\bibitem [{\citenamefont {Ng}\ \emph {et~al.}(2017)\citenamefont {Ng},
  \citenamefont {Hakim}, \citenamefont {Bhattacharjee}, \citenamefont
  {Stanier},\ and\ \citenamefont {Daughton}}]{Ng2017}%
  \BibitemOpen
  \bibfield  {author} {\bibinfo {author} {\bibfnamefont {J.}~\bibnamefont
  {Ng}}, \bibinfo {author} {\bibfnamefont {A.}~\bibnamefont {Hakim}}, \bibinfo
  {author} {\bibfnamefont {A.}~\bibnamefont {Bhattacharjee}}, \bibinfo {author}
  {\bibfnamefont {A.}~\bibnamefont {Stanier}}, \ and\ \bibinfo {author}
  {\bibfnamefont {W.}~\bibnamefont {Daughton}},\ }\bibfield  {title} {\enquote
  {\bibinfo {title} {Simulations of anti-parallel reconnection using a nonlocal
  heat flux closure},}\ }\href@noop {} {\bibfield  {journal} {\bibinfo
  {journal} {Physics of Plasmas}\ }\textbf {\bibinfo {volume} {24}},\ \bibinfo
  {pages} {082112} (\bibinfo {year} {2017})},\ \bibinfo {note}
  {{DOI:10.1063/1.4993195}}\BibitemShut {NoStop}%
\bibitem [{\citenamefont {Ng}\ \emph {et~al.}(2019)\citenamefont {Ng},
  \citenamefont {Hakim}, \citenamefont {Juno},\ and\ \citenamefont
  {Bhattacharjee}}]{Ng2019}%
  \BibitemOpen
  \bibfield  {author} {\bibinfo {author} {\bibfnamefont {J.}~\bibnamefont
  {Ng}}, \bibinfo {author} {\bibfnamefont {A.}~\bibnamefont {Hakim}}, \bibinfo
  {author} {\bibfnamefont {J.}~\bibnamefont {Juno}}, \ and\ \bibinfo {author}
  {\bibfnamefont {A.}~\bibnamefont {Bhattacharjee}},\ }\bibfield  {title}
  {\enquote {\bibinfo {title} {Drift instabilities in thin current sheets using
  a two-fluid model with pressure tensor effects},}\ }\href@noop {} {\bibfield
  {journal} {\bibinfo  {journal} {Journal of Geophysical Research: Space
  Physics}\ }\textbf {\bibinfo {volume} {124}},\ \bibinfo {pages} {3331--3346}
  (\bibinfo {year} {2019})},\ \bibinfo {note}
  {{DOI:10.1029/2018JA026313}}\BibitemShut {NoStop}%
\bibitem [{\citenamefont {Stanier}\ \emph {et~al.}(2015)\citenamefont
  {Stanier}, \citenamefont {Daughton}, \citenamefont {Chacón}, \citenamefont
  {Karimabadi}, \citenamefont {Ng}, \citenamefont {Huang}, \citenamefont
  {Hakim},\ and\ \citenamefont {Bhattacharjee}}]{Stanier2015}%
  \BibitemOpen
  \bibfield  {author} {\bibinfo {author} {\bibfnamefont {A.}~\bibnamefont
  {Stanier}}, \bibinfo {author} {\bibfnamefont {W.}~\bibnamefont {Daughton}},
  \bibinfo {author} {\bibfnamefont {L.}~\bibnamefont {Chacón}}, \bibinfo
  {author} {\bibfnamefont {H.}~\bibnamefont {Karimabadi}}, \bibinfo {author}
  {\bibfnamefont {J.}~\bibnamefont {Ng}}, \bibinfo {author} {\bibfnamefont
  {Y.~M.}\ \bibnamefont {Huang}}, \bibinfo {author} {\bibfnamefont
  {A.}~\bibnamefont {Hakim}}, \ and\ \bibinfo {author} {\bibfnamefont
  {A.}~\bibnamefont {Bhattacharjee}},\ }\bibfield  {title} {\enquote {\bibinfo
  {title} {Role of ion kinetic physics in the interaction of magnetic flux
  ropes},}\ }\href@noop {} {\bibfield  {journal} {\bibinfo  {journal} {Phys.
  Rev. Lett.}\ }\textbf {\bibinfo {volume} {115}},\ \bibinfo {pages} {175004}
  (\bibinfo {year} {2015})},\ \bibinfo {note}
  {{DOI:10.1103/PhysRevLett.115.175004}}\BibitemShut {NoStop}%
\bibitem [{\citenamefont {Powell}\ \emph {et~al.}(1999)\citenamefont {Powell},
  \citenamefont {Roe}, \citenamefont {Linde}, \citenamefont {Gombosi},\ and\
  \citenamefont {{De Zeeuw}}}]{Powell1999}%
  \BibitemOpen
  \bibfield  {author} {\bibinfo {author} {\bibfnamefont {K.}~\bibnamefont
  {Powell}}, \bibinfo {author} {\bibfnamefont {P.}~\bibnamefont {Roe}},
  \bibinfo {author} {\bibfnamefont {T.}~\bibnamefont {Linde}}, \bibinfo
  {author} {\bibfnamefont {T.}~\bibnamefont {Gombosi}}, \ and\ \bibinfo
  {author} {\bibfnamefont {D.~L.}\ \bibnamefont {{De Zeeuw}}},\ }\bibfield
  {title} {\enquote {\bibinfo {title} {A solution-adaptive upwind scheme for
  ideal magnetohydrodynamics},}\ }\href {\doibase 10.1006/jcph.1999.6299}
  {\bibfield  {journal} {\bibinfo  {journal} {J. Comp. Phys.}\ }\textbf
  {\bibinfo {volume} {154}},\ \bibinfo {pages} {284--309} (\bibinfo {year}
  {1999})}\BibitemShut {NoStop}%
\bibitem [{\citenamefont {{T{\'o}th}}\ \emph {et~al.}(2012)\citenamefont
  {{T{\'o}th}}, \citenamefont {{van der Holst}}, \citenamefont {{Sokolov}},
  \citenamefont {{De Zeeuw}}, \citenamefont {{Gombosi}}, \citenamefont
  {{Fang}}, \citenamefont {{Manchester}}, \citenamefont {{Meng}}, \citenamefont
  {{Najib}}, \citenamefont {{Powell}}, \citenamefont {{Stout}}, \citenamefont
  {{Glocer}}, \citenamefont {{Ma}},\ and\ \citenamefont {{Opher}}}]{Toth2012}%
  \BibitemOpen
  \bibfield  {author} {\bibinfo {author} {\bibfnamefont {G.}~\bibnamefont
  {{T{\'o}th}}}, \bibinfo {author} {\bibfnamefont {B.}~\bibnamefont {{van der
  Holst}}}, \bibinfo {author} {\bibfnamefont {I.~V.}\ \bibnamefont
  {{Sokolov}}}, \bibinfo {author} {\bibfnamefont {D.~L.}\ \bibnamefont {{De
  Zeeuw}}}, \bibinfo {author} {\bibfnamefont {T.~I.}\ \bibnamefont
  {{Gombosi}}}, \bibinfo {author} {\bibfnamefont {F.}~\bibnamefont {{Fang}}},
  \bibinfo {author} {\bibfnamefont {W.~B.}\ \bibnamefont {{Manchester}}},
  \bibinfo {author} {\bibfnamefont {X.}~\bibnamefont {{Meng}}}, \bibinfo
  {author} {\bibfnamefont {D.}~\bibnamefont {{Najib}}}, \bibinfo {author}
  {\bibfnamefont {K.~G.}\ \bibnamefont {{Powell}}}, \bibinfo {author}
  {\bibfnamefont {Q.~F.}\ \bibnamefont {{Stout}}}, \bibinfo {author}
  {\bibfnamefont {A.}~\bibnamefont {{Glocer}}}, \bibinfo {author}
  {\bibfnamefont {Y.-J.}\ \bibnamefont {{Ma}}}, \ and\ \bibinfo {author}
  {\bibfnamefont {M.}~\bibnamefont {{Opher}}},\ }\bibfield  {title} {\enquote
  {\bibinfo {title} {{Adaptive numerical algorithms in space weather
  modeling}},}\ }\href {\doibase 10.1016/j.jcp.2011.02.006} {\bibfield
  {journal} {\bibinfo  {journal} {Journal of Computational Physics}\ }\textbf
  {\bibinfo {volume} {231}},\ \bibinfo {pages} {870--903} (\bibinfo {year}
  {2012})}\BibitemShut {NoStop}%
\bibitem [{\citenamefont {Fonseca}\ \emph {et~al.}(2002)\citenamefont {Fonseca}
  \emph {et~al.}}]{Fonseca2002}%
  \BibitemOpen
  \bibfield  {author} {\bibinfo {author} {\bibfnamefont {R.~A.}\ \bibnamefont
  {Fonseca}} \emph {et~al.},\ }\bibfield  {title} {\enquote {\bibinfo {title}
  {Osiris: A three-dimensional, fully relativistic particle in cell code for
  modeling plasma based accelerators},}\ }\href@noop {} {\bibfield  {journal}
  {\bibinfo  {journal} {Proceedings of The International Conference on Computer
  Science}\ }\textbf {\bibinfo {volume} {2331}},\ \bibinfo {pages} {342--351}
  (\bibinfo {year} {2002})}\BibitemShut {NoStop}%
\bibitem [{\citenamefont {Hemker}(2015)}]{Hemker2015}%
  \BibitemOpen
  \bibfield  {author} {\bibinfo {author} {\bibfnamefont {R.~G.}\ \bibnamefont
  {Hemker}},\ }\bibfield  {title} {\enquote {\bibinfo {title} {Particle-in-cell
  modeling of plasma-based accelerators in two and three dimensions},}\
  }\href@noop {} {\bibfield  {journal} {\bibinfo  {journal} {arXiv:1503.00276.
  [Online]. Available: http://arxiv.org/abs/1503.00276}\ } (\bibinfo {year}
  {2015})}\BibitemShut {NoStop}%
\bibitem [{\citenamefont {Scudder}\ and\ \citenamefont
  {Daughton}(2008)}]{Scudder2008}%
  \BibitemOpen
  \bibfield  {author} {\bibinfo {author} {\bibfnamefont {J.}~\bibnamefont
  {Scudder}}\ and\ \bibinfo {author} {\bibfnamefont {W.}~\bibnamefont
  {Daughton}},\ }\bibfield  {title} {\enquote {\bibinfo {title}
  {``illuminating'' electron diffusion regions of collisionless magnetic
  reconnection using electron agyrotropy},}\ }\href@noop {} {\bibfield
  {journal} {\bibinfo  {journal} {Journal of Geophysical Research (Space
  Physics)}\ }\textbf {\bibinfo {volume} {113}},\ \bibinfo {pages} {A06222}
  (\bibinfo {year} {2008})},\ \bibinfo {note}
  {{DOI:10.1029/2008JA013035}}\BibitemShut {NoStop}%
\bibitem [{\citenamefont {Fadeev}, \citenamefont {Kvabtskhava},\ and\
  \citenamefont {Komarov}(1965)}]{Fadeev1965}%
  \BibitemOpen
  \bibfield  {author} {\bibinfo {author} {\bibfnamefont {V.~M.}\ \bibnamefont
  {Fadeev}}, \bibinfo {author} {\bibfnamefont {I.~F.}\ \bibnamefont
  {Kvabtskhava}}, \ and\ \bibinfo {author} {\bibfnamefont {N.~N.}\ \bibnamefont
  {Komarov}},\ }\bibfield  {title} {\enquote {\bibinfo {title} {An energy
  principle for hydromagnetic stability problems},}\ }\href@noop {} {\bibfield
  {journal} {\bibinfo  {journal} {Nuclear Fusion}\ }\textbf {\bibinfo {volume}
  {5}},\ \bibinfo {pages} {202} (\bibinfo {year} {1965})}\BibitemShut {NoStop}%
\bibitem [{\citenamefont {Daughton}\ \emph
  {et~al.}(2009{\natexlab{a}})\citenamefont {Daughton}, \citenamefont
  {Roytershteyn}, \citenamefont {Albright}, \citenamefont {Karimabadi},
  \citenamefont {Yin},\ and\ \citenamefont {Bowers}}]{Daughton2009_1}%
  \BibitemOpen
  \bibfield  {author} {\bibinfo {author} {\bibfnamefont {W.}~\bibnamefont
  {Daughton}}, \bibinfo {author} {\bibfnamefont {V.}~\bibnamefont
  {Roytershteyn}}, \bibinfo {author} {\bibfnamefont {B.~J.}\ \bibnamefont
  {Albright}}, \bibinfo {author} {\bibfnamefont {H.}~\bibnamefont
  {Karimabadi}}, \bibinfo {author} {\bibfnamefont {L.}~\bibnamefont {Yin}}, \
  and\ \bibinfo {author} {\bibfnamefont {K.~J.}\ \bibnamefont {Bowers}},\
  }\bibfield  {title} {\enquote {\bibinfo {title} {Influence of coulomb
  collisions on the structure of reconnection layers},}\ }\href@noop {}
  {\bibfield  {journal} {\bibinfo  {journal} {Physics of Plasmas}\ }\textbf
  {\bibinfo {volume} {16}},\ \bibinfo {pages} {072117} (\bibinfo {year}
  {2009}{\natexlab{a}})},\ \bibinfo {note}
  {{DOI:10.1063/1.3191718}}\BibitemShut {NoStop}%
\bibitem [{\citenamefont {Chen}\ and\ \citenamefont
  {T{\'o}th}(2019)}]{Chen2019}%
  \BibitemOpen
  \bibfield  {author} {\bibinfo {author} {\bibfnamefont {Y.}~\bibnamefont
  {Chen}}\ and\ \bibinfo {author} {\bibfnamefont {G.}~\bibnamefont
  {T{\'o}th}},\ }\bibfield  {title} {\enquote {\bibinfo {title} {Gauss's law
  satisfying energy-conserving semi-implicit particle-in-cell method},}\ }\href
  {\doibase https://doi.org/10.1016/j.jcp.2019.02.032} {\bibfield  {journal}
  {\bibinfo  {journal} {Journal of Computational Physics}\ }\textbf {\bibinfo
  {volume} {386}},\ \bibinfo {pages} {632--652} (\bibinfo {year}
  {2019})}\BibitemShut {NoStop}%
\bibitem [{\citenamefont {{Dong}}\ \emph {et~al.}(2021)\citenamefont {{Dong}},
  \citenamefont {{Le}}, \citenamefont {{Wang}}, \citenamefont {{Stanier}},
  \citenamefont {{Wetherton}}, \citenamefont {{Daughton}}, \citenamefont
  {{Bhattacharjee}}, \citenamefont {{Slavin}},\ and\ \citenamefont
  {{DiBraccio}}}]{Dong2021}%
  \BibitemOpen
  \bibfield  {author} {\bibinfo {author} {\bibfnamefont {C.}~\bibnamefont
  {{Dong}}}, \bibinfo {author} {\bibfnamefont {A.}~\bibnamefont {{Le}}},
  \bibinfo {author} {\bibfnamefont {L.}~\bibnamefont {{Wang}}}, \bibinfo
  {author} {\bibfnamefont {A.}~\bibnamefont {{Stanier}}}, \bibinfo {author}
  {\bibfnamefont {B.}~\bibnamefont {{Wetherton}}}, \bibinfo {author}
  {\bibfnamefont {W.}~\bibnamefont {{Daughton}}}, \bibinfo {author}
  {\bibfnamefont {A.}~\bibnamefont {{Bhattacharjee}}}, \bibinfo {author}
  {\bibfnamefont {J.}~\bibnamefont {{Slavin}}}, \ and\ \bibinfo {author}
  {\bibfnamefont {G.}~\bibnamefont {{DiBraccio}}},\ }\bibfield  {title}
  {\enquote {\bibinfo {title} {{Global Hybrid-VPIC Modeling of the Solar Wind
  Interaction with Mercury's Dynamic Magnetosphere: Reconnection and
  Foreshock}},}\ }in\ \href@noop {} {\emph {\bibinfo {booktitle} {AGU Fall
  Meeting Abstracts}}},\ Vol.\ \bibinfo {volume} {2021}\ (\bibinfo {year}
  {2021})\ pp.\ \bibinfo {pages} {P25F--2211}\BibitemShut {NoStop}%
\bibitem [{\citenamefont {{Winske}}\ \emph {et~al.}(2022)\citenamefont
  {{Winske}}, \citenamefont {{Karimabadi}}, \citenamefont {{Le}}, \citenamefont
  {{Omidi}}, \citenamefont {{Roytershteyn}},\ and\ \citenamefont
  {{Stanier}}}]{Winske2022}%
  \BibitemOpen
  \bibfield  {author} {\bibinfo {author} {\bibfnamefont {D.}~\bibnamefont
  {{Winske}}}, \bibinfo {author} {\bibfnamefont {H.}~\bibnamefont
  {{Karimabadi}}}, \bibinfo {author} {\bibfnamefont {A.}~\bibnamefont {{Le}}},
  \bibinfo {author} {\bibfnamefont {N.}~\bibnamefont {{Omidi}}}, \bibinfo
  {author} {\bibfnamefont {V.}~\bibnamefont {{Roytershteyn}}}, \ and\ \bibinfo
  {author} {\bibfnamefont {A.}~\bibnamefont {{Stanier}}},\ }\bibfield  {title}
  {\enquote {\bibinfo {title} {{Hybrid codes (massless electron fluid)}},}\
  }\href@noop {} {\bibfield  {journal} {\bibinfo  {journal} {arXiv e-prints}\
  ,\ \bibinfo {eid} {arXiv:2204.01676}} (\bibinfo {year} {2022})},\ \Eprint
  {http://arxiv.org/abs/2204.01676} {arXiv:2204.01676 [physics.plasm-ph]}
  \BibitemShut {NoStop}%
\bibitem [{\citenamefont {{Makwana}}, \citenamefont {{Keppens}},\ and\
  \citenamefont {{Lapenta}}(2018)}]{Makwana2018}%
  \BibitemOpen
  \bibfield  {author} {\bibinfo {author} {\bibfnamefont {K.}~\bibnamefont
  {{Makwana}}}, \bibinfo {author} {\bibfnamefont {R.}~\bibnamefont
  {{Keppens}}}, \ and\ \bibinfo {author} {\bibfnamefont {G.}~\bibnamefont
  {{Lapenta}}},\ }\bibfield  {title} {\enquote {\bibinfo {title} {{Two-way
  coupled MHD-PIC simulations of magnetic reconnection in magnetic island
  coalescence}},}\ }in\ \href {\doibase 10.1088/1742-6596/1031/1/012019} {\emph
  {\bibinfo {booktitle} {Journal of Physics Conference Series}}},\ \bibinfo
  {series} {Journal of Physics Conference Series}, Vol.\ \bibinfo {volume}
  {1031}\ (\bibinfo {year} {2018})\ p.\ \bibinfo {pages} {012019}\BibitemShut
  {NoStop}%
\bibitem [{\citenamefont {Karimabadi}\ \emph {et~al.}(2011)\citenamefont
  {Karimabadi}, \citenamefont {Dorelli}, \citenamefont {Roytershteyn},
  \citenamefont {Daughton},\ and\ \citenamefont {Chacón}}]{Karimabadi2011_1}%
  \BibitemOpen
  \bibfield  {author} {\bibinfo {author} {\bibfnamefont {H.}~\bibnamefont
  {Karimabadi}}, \bibinfo {author} {\bibfnamefont {J.}~\bibnamefont {Dorelli}},
  \bibinfo {author} {\bibfnamefont {V.}~\bibnamefont {Roytershteyn}}, \bibinfo
  {author} {\bibfnamefont {W.}~\bibnamefont {Daughton}}, \ and\ \bibinfo
  {author} {\bibfnamefont {L.}~\bibnamefont {Chacón}},\ }\bibfield  {title}
  {\enquote {\bibinfo {title} {Flux pileup in collisionless magnetic
  reconnection: Bursty interaction of large flux ropes},}\ }\href@noop {}
  {\bibfield  {journal} {\bibinfo  {journal} {Physical Review Letters}\
  }\textbf {\bibinfo {volume} {107}},\ \bibinfo {pages} {025002} (\bibinfo
  {year} {2011})},\ \bibinfo {note}
  {{DOI:10.1103/PhysRevLett.107.025002}}\BibitemShut {NoStop}%
\bibitem [{\citenamefont {Daughton}\ \emph
  {et~al.}(2009{\natexlab{b}})\citenamefont {Daughton}, \citenamefont
  {Roytershteyn}, \citenamefont {Albright}, \citenamefont {Karimabadi},
  \citenamefont {Yin},\ and\ \citenamefont {Bowers}}]{Daughton2009_2}%
  \BibitemOpen
  \bibfield  {author} {\bibinfo {author} {\bibfnamefont {W.}~\bibnamefont
  {Daughton}}, \bibinfo {author} {\bibfnamefont {V.}~\bibnamefont
  {Roytershteyn}}, \bibinfo {author} {\bibfnamefont {B.~J.}\ \bibnamefont
  {Albright}}, \bibinfo {author} {\bibfnamefont {H.}~\bibnamefont
  {Karimabadi}}, \bibinfo {author} {\bibfnamefont {L.}~\bibnamefont {Yin}}, \
  and\ \bibinfo {author} {\bibfnamefont {K.~J.}\ \bibnamefont {Bowers}},\
  }\bibfield  {title} {\enquote {\bibinfo {title} {Transition from collisional
  to kinetic regimes in large-scale reconnection layers},}\ }\href@noop {}
  {\bibfield  {journal} {\bibinfo  {journal} {Physical Review Letters}\
  }\textbf {\bibinfo {volume} {103}},\ \bibinfo {pages} {065004} (\bibinfo
  {year} {2009}{\natexlab{b}})},\ \bibinfo {note}
  {{DOI:10.1103/PhysRevLett.103.065004}}\BibitemShut {NoStop}%
\bibitem [{\citenamefont {{Ji}}\ \emph {et~al.}(2020)\citenamefont {{Ji}},
  \citenamefont {{Yoo}}, \citenamefont {{Jara-Almonte}}, \citenamefont
  {{Goodman}}, \citenamefont {{Ren}}, \citenamefont {{Yamada}}, \citenamefont
  {{Bhattacharjee}}, \citenamefont {{Fox}}, \citenamefont {{Daughton}},
  \citenamefont {{Stanier}},\ and\ \citenamefont {{The Flare Construction Team
  Team}}}]{Ji2020}%
  \BibitemOpen
  \bibfield  {author} {\bibinfo {author} {\bibfnamefont {H.}~\bibnamefont
  {{Ji}}}, \bibinfo {author} {\bibfnamefont {J.}~\bibnamefont {{Yoo}}},
  \bibinfo {author} {\bibfnamefont {J.}~\bibnamefont {{Jara-Almonte}}},
  \bibinfo {author} {\bibfnamefont {A.}~\bibnamefont {{Goodman}}}, \bibinfo
  {author} {\bibfnamefont {Y.}~\bibnamefont {{Ren}}}, \bibinfo {author}
  {\bibfnamefont {M.}~\bibnamefont {{Yamada}}}, \bibinfo {author}
  {\bibfnamefont {A.}~\bibnamefont {{Bhattacharjee}}}, \bibinfo {author}
  {\bibfnamefont {W.}~\bibnamefont {{Fox}}}, \bibinfo {author} {\bibfnamefont
  {W.}~\bibnamefont {{Daughton}}}, \bibinfo {author} {\bibfnamefont
  {A.}~\bibnamefont {{Stanier}}}, \ and\ \bibinfo {author} {\bibnamefont {{The
  Flare Construction Team Team}}},\ }\bibfield  {title} {\enquote {\bibinfo
  {title} {{FLARE: a collaborative research facility to study magnetic
  reconnection and related phenomena}},}\ }in\ \href@noop {} {\emph {\bibinfo
  {booktitle} {APS Division of Plasma Physics Meeting Abstracts}}},\ \bibinfo
  {series} {APS Meeting Abstracts}, Vol.\ \bibinfo {volume} {2020}\ (\bibinfo
  {year} {2020})\ p.\ \bibinfo {pages} {NO04.008}\BibitemShut {NoStop}%
\bibitem [{\citenamefont {Ebrahimi}\ and\ \citenamefont
  {Raman}(2016)}]{Ebrahimi2016}%
  \BibitemOpen
  \bibfield  {author} {\bibinfo {author} {\bibfnamefont {F.}~\bibnamefont
  {Ebrahimi}}\ and\ \bibinfo {author} {\bibfnamefont {R.}~\bibnamefont
  {Raman}},\ }\bibfield  {title} {\enquote {\bibinfo {title} {Large-volume flux
  closure during plasmoid-mediated reconnection in coaxial helicity
  injection},}\ }\href {\doibase 10.1088/0029-5515/56/4/044002} {\bibfield
  {journal} {\bibinfo  {journal} {Nuclear Fusion}\ }\textbf {\bibinfo {volume}
  {56}},\ \bibinfo {pages} {044002} (\bibinfo {year} {2016})}\BibitemShut
  {NoStop}%
\bibitem [{\citenamefont {Tanabe}\ \emph {et~al.}(2017)\citenamefont {Tanabe},
  \citenamefont {Yamada}, \citenamefont {Watanabe}, \citenamefont {Gi},
  \citenamefont {Inomoto}, \citenamefont {Imazawa}, \citenamefont
  {Gryaznevich}, \citenamefont {Michael}, \citenamefont {Crowley},
  \citenamefont {Conway}, \citenamefont {Scannell}, \citenamefont {Harrison},
  \citenamefont {Fitzgerald}, \citenamefont {Meakins}, \citenamefont {Hawkes},
  \citenamefont {McClements}, \citenamefont {O'Gorman}, \citenamefont {Cheng},\
  and\ \citenamefont {Ono}}]{Tanabe2017}%
  \BibitemOpen
  \bibfield  {author} {\bibinfo {author} {\bibfnamefont {H.}~\bibnamefont
  {Tanabe}}, \bibinfo {author} {\bibfnamefont {T.}~\bibnamefont {Yamada}},
  \bibinfo {author} {\bibfnamefont {T.}~\bibnamefont {Watanabe}}, \bibinfo
  {author} {\bibfnamefont {K.}~\bibnamefont {Gi}}, \bibinfo {author}
  {\bibfnamefont {M.}~\bibnamefont {Inomoto}}, \bibinfo {author} {\bibfnamefont
  {R.}~\bibnamefont {Imazawa}}, \bibinfo {author} {\bibfnamefont
  {M.}~\bibnamefont {Gryaznevich}}, \bibinfo {author} {\bibfnamefont
  {C.}~\bibnamefont {Michael}}, \bibinfo {author} {\bibfnamefont
  {B.}~\bibnamefont {Crowley}}, \bibinfo {author} {\bibfnamefont {N.~J.}\
  \bibnamefont {Conway}}, \bibinfo {author} {\bibfnamefont {R.}~\bibnamefont
  {Scannell}}, \bibinfo {author} {\bibfnamefont {J.}~\bibnamefont {Harrison}},
  \bibinfo {author} {\bibfnamefont {I.}~\bibnamefont {Fitzgerald}}, \bibinfo
  {author} {\bibfnamefont {A.}~\bibnamefont {Meakins}}, \bibinfo {author}
  {\bibfnamefont {N.}~\bibnamefont {Hawkes}}, \bibinfo {author} {\bibfnamefont
  {K.~G.}\ \bibnamefont {McClements}}, \bibinfo {author} {\bibfnamefont
  {T.}~\bibnamefont {O'Gorman}}, \bibinfo {author} {\bibfnamefont {C.~Z.}\
  \bibnamefont {Cheng}}, \ and\ \bibinfo {author} {\bibfnamefont
  {Y.}~\bibnamefont {Ono}},\ }\bibfield  {title} {\enquote {\bibinfo {title}
  {Recent progress of magnetic reconnection research in the mast spherical
  tokamak},}\ }\href {\doibase 10.1063/1.4977922} {\bibfield  {journal}
  {\bibinfo  {journal} {Physics of Plasmas}\ }\textbf {\bibinfo {volume}
  {24}},\ \bibinfo {pages} {056108} (\bibinfo {year} {2017})}\BibitemShut
  {NoStop}%
\bibitem [{\citenamefont {{Olson}}\ \emph {et~al.}(2021)\citenamefont
  {{Olson}}, \citenamefont {{Egedal}}, \citenamefont {{Clark}}, \citenamefont
  {{Endrizzi}}, \citenamefont {{Greess}}, \citenamefont {{Millet-Ayala}},
  \citenamefont {{Myers}}, \citenamefont {{Peterson}}, \citenamefont
  {{Wallace}},\ and\ \citenamefont {{Forest}}}]{Olson2021}%
  \BibitemOpen
  \bibfield  {author} {\bibinfo {author} {\bibfnamefont {J.}~\bibnamefont
  {{Olson}}}, \bibinfo {author} {\bibfnamefont {J.}~\bibnamefont {{Egedal}}},
  \bibinfo {author} {\bibfnamefont {M.}~\bibnamefont {{Clark}}}, \bibinfo
  {author} {\bibfnamefont {D.~A.}\ \bibnamefont {{Endrizzi}}}, \bibinfo
  {author} {\bibfnamefont {S.}~\bibnamefont {{Greess}}}, \bibinfo {author}
  {\bibfnamefont {A.}~\bibnamefont {{Millet-Ayala}}}, \bibinfo {author}
  {\bibfnamefont {R.}~\bibnamefont {{Myers}}}, \bibinfo {author} {\bibfnamefont
  {E.~E.}\ \bibnamefont {{Peterson}}}, \bibinfo {author} {\bibfnamefont
  {J.}~\bibnamefont {{Wallace}}}, \ and\ \bibinfo {author} {\bibfnamefont
  {C.~B.}\ \bibnamefont {{Forest}}},\ }\bibfield  {title} {\enquote {\bibinfo
  {title} {{Regulation of the normalized rate of driven magnetic reconnection
  through shocked flux pileup}},}\ }\href {\doibase 10.1017/S0022377821000659}
  {\bibfield  {journal} {\bibinfo  {journal} {Journal of Plasma Physics}\
  }\textbf {\bibinfo {volume} {87}},\ \bibinfo {eid} {175870301} (\bibinfo
  {year} {2021})}\BibitemShut {NoStop}%
\end{thebibliography}

\providecommand{\noopsort}[1]{}\providecommand{\singleletter}[1]{#1}%

\end{document}